\documentstyle[pre,aps,floats,graphics,epsfig]{revtex}
\newcommand{\be}{\begin{equation}}
\newcommand{\ee}{\end{equation}}
\newcommand{\bea}{\begin{eqnarray}}
\newcommand{\eea}{\end{eqnarray}}

\newcommand{\beet}{\begin{equation*}}
\newcommand{\eeet}{\end{equation*}}
\newcommand{\beaet}{\begin{eqnarray*}}
\newcommand{\eeaet}{\end{eqnarray*}}
\newcommand{\bfig}{\begin{figure}}
\newcommand{\efig}{\end{figure}}
\newcommand{\bc}{\begin{center}}
\newcommand{\ec}{\end{center}}

\newcommand{\szz}{\sigma_{tt}}
\newcommand{\sxx}{\sigma_{xx}}
\newcommand{\szx}{\sigma_{tx}}
\newcommand{\sxz}{\sigma_{xt}}
\newcommand{\snn}{\sigma_{nn}}

\newcommand{\sns}{\sigma_{nm}}
\newcommand{\sij}{\sigma_{ij}}
\newcommand{\sab}{\sigma_{\alpha\beta}}

\newcommand{\iqt}{\int_{-\infty}^{+\infty} \quad {\rm d}q \,}

\newcommand{\de}{\delta}
\newcommand{\La}{\Lambda}
\newcommand{\ga}{\gamma}
\newcommand{\gah}{\hat{\gamma}}

\newcommand{\ra}{\right >}
\newcommand{\la}{\left <}

\newcommand{\dr}{\partial}

\newcommand{\dy}{\raisebox{1.6ex}{\rotatebox{180}{\textsf{Y}}}}
\newcommand{\uy}{\textsf{Y}}

\begin{document}

\title{Granular media: some ideas from statistical physics}

\author{J.P. Bouchaud}
\address{
Service de Physique de l'Etat Condens\'e,\\ CEA, Orme des
Merisiers,\\ 91191 Gif-sur-Yvette, Cedex France. }
\maketitle
\small{These lecture notes cover the statics and 
glassy dynamics of granular media. Most of the lectures were in fact devoted
to `force propagation' models. We discuss the experimental and theoretical 
motivations for these approaches, and their conceptual connections with Edwards'
thermodynamical analogy. One of the distinctive feature of granular media (common to
many other `jammed' systems) is indeed 
the large number of metastable states that are macroscopically
equivalent. We present in detail the (scalar) $q$-model and its
tensorial generalization, that aim at modelling the existence of force chains and
arching effects without introducing any displacement field. The contrast between 
the hyperbolic equations obtained within this line
of thought and elliptic (elastic) equations is emphasized. The r\^ole of disorder 
on these hyperbolic
equations is studied in details using perturbative and diagrammatic methods. 
Recent (strong disorder) force chain network models are reviewed, and 
compared with the experimental determination of the force `response function' in granular materials.
We briefly discuss several issues (such as isostaticity and generic marginality) and open problems.
At the end of these notes, we also discuss the basic dynamical properties of 
{\it weakly tapped} 
granular assemblies, and stress the phenomenological analogies with other glassy materials.
Simple models that account for slow compaction and dynamical heterogeneities are presented, 
that are inspired by `free-volume' ideas and Edwards' assumption. A connection with the
theory of fluctuating random surfaces, also noted recently by Castillo et al., is 
suggested. Finally, we discuss how the `trap
model' can be adapted to granular materials, such that more subtle `memory' effects can be accounted
for.}
 
\section{Introduction}

\subsection{Basic phenomenology}

Although granular materials are made of classical particles of macroscopic size,
they exhibit a host of interesting and sometimes counter-intuitive properties, which
are of interest both to the academic community and for industrial applications: 
enormous amounts of `granular assemblies' are routinely handled (stored, transported, mixed
together, etc.) The reason why these systems are still not fully 
understood yet is that they 
require a proper statistical treatment of collective effects. Although the physics 
at the grain level is reasonably well understood, the behaviour of a large assembly
of these grains, with strongly non-linear interactions, demand concepts and methods
that only recently emerged from the study of disordered systems in statistical mechanics.
The main property of granular materials, that leads both to most of the non trivial 
phenomenology and to most of the theoretical difficulties, is the existence of 
a {\it large number of different microscopic metastable states} that are macroscopically 
equivalent. It turns out, quite interestingly, that this feature (metastability) is common to a 
wider class of materials in their `jammed' state. This includes glasses, colloids, 
compressed emulsions and foams, spin-glasses, vortex glasses and other collectively 
pinned structures, etc.

There is a vast body of experimental results on granular materials that we do not aim 
to cover here (for reviews, see \cite{RJaeger,RCargese,RDuran,RCRAS}). We will mainly 
restrict to {\it dry} grains in static and weakly driven 
situations. Correspondingly, we will focus on the statistical properties of the {\it static 
states} (and
in particular the distribution of stresses) and on the glassy dynamics of {\it gently 
`tapped' assemblies}, that slowly evolve from one static state to another. We have therefore
excluded from these lectures, because of lack of time, `strongly' driven situations, 
such as granular flows, avalanches and surface flows, dune formation, etc. This is not to
suggest that these problems are less interesting and that collective effects are 
irrelevant. Quite on the contrary, the dynamics of strongly driven granular systems 
also displays remarkable effects, such as collisional clustering which generates non trivial
spatial structures in granular flows, and invalidates simple hydrodynamical descriptions 
\cite{GoldCargese,Luding,BonamyPRL}.
Some recent papers on these matters can be found in references \cite{RCargese,RCRAS,PoechelBook,Poechel,Poechel2}.

Stress patterns in dry granular media exhibit some rather unusual features when
compared to either liquids or elastic solids. For example, the
vertical pressure below conical sand-piles does not follow the height of
material above a particular point. Depending on the way the pile
is prepared, it shows a {\it minimum} underneath
the apex of the pile \cite{Smid,Huntley,Vaneltas} when the pile is built from a
point source, and a broad, flat maximum when it is built layer by layer from 
a uniform `rain' of grains. Furthermore, local stress
fluctuations are large, sometimes on length scales much larger than the
grain size. For
example, repeatedly pouring the very same amount of powder in a silo results in
fluctuations of the weight supported by the bottom plate of $20 \%$ or more
\cite{Brown,Vanelsilo}. Weak perturbations of the packing can sometimes cause 
large rearrangements \cite{RouxPRL,Dauchot}. Qualitatively, these features are attributed to 
the presence of {\it stress paths} which can focus the stress field into localized
regions and also deflect it to cause ``arching" (see \cite{Dantuetc} for early qualitative 
experiments). More quantitative experiments were performed in \cite{Liu,Huntley,fluctuations}, 
where the local
fluctuations of the normal stress deep inside a silo or at the base of a
sandpile were measured. It was found that the stress probability distribution is rather
broad, decaying exponentially for large stresses. This behaviour was also
found in numerical simulations \cite{Radjai}, and more recently in other situations \cite{NagelG}, such as
compressed emulsions \cite{Camb}. Similarly, standard `triaxial' test experiments used
to determine the elastic properties of materials from the (macroscopic) relation between
stresses and deformations show highly irreproducible, hysteretic behaviour which only seem to 
converge towards a well defined curve after a large number of deformation cycles have
been imposed to the granular system in order to `anneal' it down to a reproducible state.

The dynamics of slowly driven granular systems also exhibit unusual features when
compared to either liquids or solids, and has actually much in common with {\it glasses}.
In particular, the way these systems very slowly compact when vibrated, the unusual 
dependence of the density on the system history, etc. has strong similarities with 
the properties of glassy materials. At the phenomenological level, the dynamics of 
these systems appears to be a succession of hops between different (metastable) 
equilibrium states. The understanding of static and weakly driven granular assemblies 
therefore require a proper description of the statistical properties of these `blocked'
configurations. We now discuss these issues in the perspective of the present lectures.
\footnote{The
content of these lecture notes owes a lot to my collaborators on these issues: Mike Cates, 
Philippe
Claudin, Eric Cl\'ement, Dov Levine, Josh Socolar, Matthias Otto and Joachim Wittmer. Parts of these notes actually are 
extracted
from various papers co-written with them. I have tried to add my own present understanding 
of the subject,
in particular concerning Edwards ensembles, the hyperbolic approach to the statics of granular 
media, and their glassy dynamics, in particular dynamical heterogeneities. Some ideas are still speculative and are by no means intended to be
definitive, but I hope that 
the theoretical concepts and methods are of sufficiently broad interest to deserve appearing 
in print in the present Les Houches volume. I wish to express my gratitude to Jean-Louis Barrat 
and Jorge Kurchan 
for giving me the 
opportunity of giving these lectures and for many very inspiring discussions. 
I thank the participants of the school 
for interesting comments and ideas, in particular G. Biroli, L. Berthier,
E. Bertin, L. Cugliandolo, D. Fisher, A. Lefevre, 
J. Snoeijer, V. Viasnoff and O. White. I also thank E. Bertin for carefully reading the manuscript, and D. Bonamy,
O. Dauchot, F. Daviaud, C. Godr\`eche, M. M\'ezard, J. N. Roux, R. da Silveira and E. Vincent 
for discussions.} 

\subsection{Theoretical issues}

\subsubsection{Static properties}

How can one then describe the statics of granular materials on large length
scales? The basic problem stems from the fact that the equilibrium equations
for the stress tensor are not sufficient to determine the stress. For example, 
in two dimensions the stress tensor has three independent components, but there
are only two equilibrium equations. Some
additional assumptions about the properties of the material must be provided. 
For example, the assumption that the material is elastic and follows Hooke's law
gives extra constraints on the stress tensor and allows one to solve the static
problem as soon as some appropriate boundary conditions are given. For 
granular materials, the standard procedure is to use elastic or elasto-plastic
theories from soil mechanics \cite{Wood}. However, the relation between
force chains on short length scales and an elasto-plastic description on
large length scales is far from obvious. To our knowledge, no systematic procedure
has ever been proposed to go from the mechanics at the grain level to coarse-grained
equations that would justify the use of an elasto-plastic framework and estimate the
parameters of the theory (effective elastic moduli, etc.). One of the main difficulty 
is that some indeterminacy exists already at the grain level, since many different 
configurations of the contact forces are allowed and satisfy local equilibrium. 
This leads to several conceptual
problems: even if an elastic-like description of small perturbations around 
an arbitrary reference state (such as sound waves, for example) might make 
sense in general, the description of -- say -- a conical sandpile using 
elasto-plasticity theory \cite{Cantelaube} requires the identification of a (zero stress) reference state 
from which deformations can be defined, at least for {\it some} regions of the pile. 
In the case of a pile of infinitely hard grains 
(which should be the correct benchmark for an assembly of grains with a Young modulus 
much larger than the gravity induced stresses) that rests in one particular metastable
state (among a large number of macroscopically equivalent ones), switching
the gravity back to zero will hardly affect the packing. Each of these
metastable states can thus equally well be taken as a reference state; 
on the other hand, it is precisely the stress pattern in one of these `native' metastable state 
(i.e. obtained
when grains come to rest without further tapping) that one wants to predict. One peculiarity 
of (dry) granular 
materials is the absence
of tensile stresses between the grains; the cohesion of the assembly is therefore induced by
the applied stress itself and the zero stress state is ill defined.

Even the description of small perturbations around a given reference state 
might be problematic. For example, the existence of a (large volume) limiting curve 
relating incremental stresses
and deformations requires, as already mentioned above, at least some `annealing' procedure
to define a reproducible initial state. This 
limiting curve might not even exist in the absence of friction 
\cite{RouxPRL}. Even for moderate deformations, following the so-called consolidation
phase, the response to cyclic loads in standard triaxial tests shows some
significant irreversibility. 

The absence of any obvious deformation field from which the stress tensor
may be constructed has motivated an alternative, `stress-only' approach \cite{BCC,WCCB,us,these}. 
The basic tenet of these theories is that in
equilibrium, some (history dependent) large scale relations between the
components of the stress tensor should exist. These relations should be determined 
by the global statistical features of the particular metastable
state in which the packing sits but not on its microscopic details, nor on the particular
loading conditions, provided these do not lead to further rearrangements of the packing. 
Much as random collisions between molecules give rise, on large
length scale, to well defined hydrodynamical equations, the hope is that an appropriate
coarse-graining of the local force balance equations leads, on large length scales, 
to the missing `closure' equation that allows to solve for the static equilibrium. (For 
rather formal attempts in this direction, see \cite{Grinev,Ball}.)
A well known relation of this type arises from the assumption that the
material is everywhere on the verge of plastic failure, leading to a Mohr-Coulomb
(non-linear) relation between the stress components \cite{Nedderman}, but we will
motivate and discuss simpler relations below, based both on symmetry arguments 
\cite{BCC} and on the consideration of simple rules for the transfer of stresses between 
adjacent grains \cite{pre}.  The consequence of a fixed relation
between the components of the stress tensor is that stresses obey an 
\textit{hyperbolic} equation, as
compared to the \textit{elliptic} equations encountered in elasticity theory. This
means that stresses `propagate' or are `transmitted' along lines: as discussed below, the
characteristics of this hyperbolic equation are the
mathematical transcription of the force chains that are well known
to exist in granular materials \cite{prl}.

\subsubsection{Tapping and non thermal ensembles}
\label{halsey}
As mentioned above, many different packings and 
configurations of the contact forces are compatible with the local equilibrium of each grain
for a given macroscopic situation. This is actually intimately related to the fact that stresses 
in granular media often show large fluctuations; some
kind of averaging is therefore needed to obtain reproducible results. In the case of 
sand-piles, one must repeat the construction of the pile several times, and use a pressure 
gauge that averages over a sufficiently large number of grains, in order to obtain a satisfactory 
stress profile that a statistical theory of blocked states should predict. Another 
possibility is to vibrate the packing such as to make it probe, during its evolution, 
several equilibrium states with the same macroscopic geometry. The natural 
question is then: with which statistical weight the different equilibrium (blocked) states 
appear in a given experiment? To what extent are these weights dependent on the dynamics 
that leads to the blocked states? Is the ensemble of `native' (as-built) packings identical to
the ensemble of packings obtained under tapping? 

The simplest answer, proposed more than ten years ago 
by Sam Edwards, is to postulate that all blocked states with a given density are equiprobable \cite{EdwardsE}. 
This micro-canonical assumption defines what is now called the `Edwards ensemble' \cite{Kurchan}; it turns out 
that several 
toy models of jammed systems do obey, either exactly or to a good approximation, 
Edwards' prescription \cite{Berg,Lefevre,Godreche,Coniglio}. 
This is a first step towards a `thermodynamical' description of 
out-of-equilibrium, dissipative systems \cite{Kurchan2,Maakse,BerthierBarrat,ABarrat}. 
However, several remarks of various nature should be made here. 

\begin{itemize}

\item First, the analogy between tapping strength and temperature. In many cases, this is
a useful intuitive guide and several experiments discussed in section ~\ref{dynamics} do indeed 
confirm the phenomenological analogies between the two. However, tapping is a long-wavelength 
excitation, whereas temperature in solid state physics is thought to give rise to very
short wavelength fluctuations. Although the long-wavelength excitation probably cascades 
down, through collisions, to short wavelengths, ideas such as detailed balance and activated 
processes might be affected by the correlated nature of the noise. In this respect, the 
non trivial clustering patterns induced by the dissipative collisions might also obliterate 
simple ideas on the statistics of blocked states. 

\item One must distinguish at least two types of tapping excitations. One would be very 
gentle taps, that are insufficient to change the packing {\it geometry}, but do change 
the contact forces for each grain. In this case, tapping induces a random walk in `force 
space', but for a fixed configuration of the grains. The Edwards hypothesis in this restricted case
is to assign a uniform weight for all force configurations that (i) lead to static
equilibrium (forces and torques on each grain add up to zero) and (ii) satisfy the Coulomb 
inequality at each contact. One can also drive the system with an amplitude such that 
the motion of grains is possible, in which case the dynamics is a random walk both in 
force space and in packing space. The extended Edwards ensemble in this case is to assign 
equal weight to any packing and any force configuration such that equilibrium is obeyed 
and the Coulomb inequality satisfied. In principle, the Edwards prescription could be
correct in one case and not in the other, or in both, or in none, or more complicated situations
still. 

\item The Edwards prescription is however ambiguous for continuous variables. In the `gentle'
tap case, one is tempted to interpret Edwards' measure as follows. Let us call 
$\vec f_i^\alpha$ the contact force on the $\alpha$-th contact of the $i$-th grain, 
and $\vec r_i^\alpha$ the position of the contact point. We call $\mu$ the friction 
coefficient, and the indices $N$ and $T$ refer to the normal and tangential 
components of the force. The natural Edwards measure reads:
\be
P\left(\{\vec f_i^\alpha\}\right)= \frac{1}{Z} \prod_{i} \left[\delta\left(\sum_\alpha \vec f_i^\alpha\right)
\delta\left(\sum_\alpha \vec f_i^\alpha \times \vec r_i^\alpha \right)
\prod_{\alpha} \Theta\left(\mu f_{i,N}^\alpha - |f_{i,T}^\alpha| \right)\right],
\ee
which is a formal way to impose the constraints on each grain. ($\Theta$ is the step function). 
However, this assumes that
the {\it a priori} measure on the forces is uniform, which is reasonable but not obvious. 
The usual microcanonical ensemble for particles is constructed similarly: one imposes the
total energy of the system using a $\delta$-function on an {\it a priori} uniform measure
on the canonical variables (position and momentum). However, in the latter case, this procedure
is justified by the Liouville theorem which selects the relevant canonical variables. In
general, however, there is an ambiguity since the assumption of uniformity is not invariant 
under changes of variables. 

\end{itemize}

It is instructive to discuss the simplest case where the Edwards assumption can be discussed
in details, and perhaps tested experimentally or numerically. Consider, as proposed by 
Ertas and Halsey, a single disk in a wedge \cite{Ertas}. In equilibrium, there are two contact points
and therefore four unknowns: $f_{1,N},f_{1,T}$ and $f_{2,N},f_{2,T}$, where $f_{\alpha,T} > 0$
means that the force pushes upwards. These forces must 
lead to equilibrium, which gives three equations. There is therefore one degree of 
freedom which is not fixed by the equilibrium requirement, and is dynamically selected. 
It is easy to see that one must have $f_{1,T}=f_{2,T}=f_T$ and $f_{1,N}=f_{2,N}=f_N$. 
The Edwards measure then reads:
\be
P(f_N,f_T) = \frac{1}{Z}\, \delta\left(f_N \sin \psi + f_T \cos \psi - \frac12 Mg\right) 
\Theta(\mu f_N - |f_T|),
\ee
where $\psi$ is (half) the opening angle of the wedge. From this result, 
one can compute the distribution of the `mobilization' ratio $r = f_T/f_N$, which is 
found to be parabolic: 
\be
P(r) \propto \left(\sin \psi+r \cos \psi\right)^2 \qquad (-\mu \leq r \leq \mu),
\ee
and, of course, zero outside the allowed interval $[-\mu,\mu]$. One could test this simple 
predictions by repeatedly tapping spheres made of different materials, and investigate 
the relevance of the tapping mode and the contact dynamics on the statistical ensemble of 
forces that one generates. This would be quite a valuable starting point, before speculating 
on more complex multi-grain situations. We will discuss below some numerical results that 
indeed suggest some dependence of the statistics of forces on the microscopic dynamics. 

The Edwards prescription is in fact at the heart of the simplest `scalar' model for the 
statistics of forces in granular materials, which was proposed in \cite{Liu,Copper}
to account for the empirical exponential tail in the distribution of forces. 
Although this model represents a highly stylized view of granular systems 
and cannot be expected to be accurate, it provides both an extremely rich theoretical
benchmark and a pedagogical starting point for more elaborate descriptions.

\section{The Scalar Model I: Discrete version}
\label{ScalarI}

\subsection{Definition and motivation}

The drastic simplification of the scalar model is to only retain one component 
of the stress tensor, namely the `weight' $w = \sigma_{zz}$, and correspondingly, to
only consider the force balance equation along the vertical axis. Again for simplicity,
one can think that the grains reside on the nodes of a two-dimensional lattice, and
are labeled by two integers: $i$ in the horizontal direction and $j$ in the vertical 
direction; $j$ increases as one moves downwards. The equilibrium equation can then be 
written as:
\be
\label{liudiscret}
w(i,j) = w_0 + q_+(i-1,j-1)w(i-1,j-1) + q_-(i+1,j-1)w(i+1,j-1)
\ee
where `$w_0$' is the weight of each grain, and $q_\pm(i,j)$ are `transmission' 
coefficients giving the fraction of weight which the $(i,j)$ transmits to its 
right (resp. left) neighbour immediately below, such that $q_+(i,j) + q_-(i,j)=1$ for all 
$i,j$'s. 
The case of an ordered pile of identical grains and identical conditions for each contacts 
corresponds to 
$q_\pm = \frac{1}{2}$.
In this case, the equation for the $w$'s become identical to the Master equation
describing the population of un-biased random walkers in a one space dimension 
(corresponding to $i$), evolving in `time' (corresponding to $j$):
\be
\label{rw}
w(i,j) = w_0 + \frac{1}{2} \left[w(i-1,j-1) + w(i+1,j-1)\right].
\ee
The term $w_0$ is a constant source of particles that are created uniformly in 
space and in time. We will explore this analogy further below.

Now, grain assemblies are usually not perfectly ordered: grains have various shapes 
and sizes; there are packing defects and irregularities; even for a perfectly 
ordered packing of identical grains one can expect that the history has imposed 
different contact loadings. In the above language, it means that the $q_\pm(i,j)$ are
not all identical and reflect the above sources of randomness. The idea of Edwards,
in this highly simplified framework, can easily be worked out, since each 
`blocked' state corresponds to a particular choice of -- say -- $q_+(i,j)$ for 
all $i,j$. Provided $q_-(i,j)=1 -q_+(i,j)$, equilibrium is ensured. The uniform
measure on all blocked states, advocated by Edwards, merely translates, in the present case, 
as a uniform probability distribution for $q_+$ (or $q_-$) between $0$ and $1$. 
This defines the $q$ model, which was originally written with an arbitrary number $N$
of downward neighbours ($N=2$ in the example above), and can thus be (in principle)
generalized to three dimensions \cite{Liu,Copper}. In this case, there are $N$ 
coefficients $q_\alpha$, $\alpha=1,\cdots, N$ per grain, and the Edwards measure 
corresponds to the choosing all the $q_\alpha$ independently on each node $i,j$ such that:
\be
P(\{q_\alpha\}) = \frac{1}{Z} \delta\left(\sum_{\alpha=1}^N q_\alpha - 1\right).
\ee 
Therefore, in the present case, Edwards prescription can be explicitly followed. 
It may seem {\it a priori} that this microcanonical assumption is too simple and can only 
lead to trivial predictions. However, as we discuss below, this is not the case:
much as for gases where the microcanonical hypothesis can be used to derive, for
example, the Maxwell distribution for the particle velocities, the $q$-model predicts
a non trivial distribution for the local vertical stress. 

\subsection{Stress distribution and the exponential tail}

The case of a uniform distribution of the $q$'s is interesting because it 
leads to an exact solution for the local weight distribution $P(w)$ for large
heights. Let us assume for the moment that the weights on neighbouring sites 
become asymptotically independent (\cite{Copper,Jacco}). Then $P(w)$ obeys the following mean-field 
equation:
\be
P_{j+1}(w) = \int_0^1 dq_1 dq_2 \rho(q_1)\rho(q_2) \int_0^\infty dw_1 dw_2
P_{j}(w_1)P_{j}(w_2) \delta[w-(w_1 q_1 + w_2 q_2 +w_0)] \label{recu}
\ee
where $\rho(q)$ is the distribution of $q$, here taken to be $\rho(q) = 1$.
In the limit $j \to \infty$, the stationary distribution $P^*$ of this equation
can be explicitly constructed and is given by:
\be
P^*(w) = \frac{w}{\overline{w}^2} \exp-\frac{w}{\overline{w}}\label{exptail}
\ee
where $2 \overline{w}= j w_0$ is the average weight. For $N \neq 2$, the distribution 
is instead a Gamma distribution of parameter $N$; its small $w$ behaviour is $w^{N-1}$
while the large $w$ tail is exponential. Liu et al. \cite{Liu,Copper} have argued that
this behaviour is generic and survives deviations away from the strict Edwards 
prescription: for example, the condition for the local weight $w$ to be
small is that all the $N$ $q$'s reaching this site 
are themselves small; the phase space volume for this is proportional to $w^{N-1}$
if the distribution $\rho(q)$ is regular around $q=0$. However, if instead $\rho(q)
\propto q^{\gamma-1}$ when $q$ is small, one expects $P^*(w)$ to behave for small $w$ 
as $w^{-\alpha}$, with $\alpha=1-N\gamma < 0$. 

Similarly, the exponential tail at large $w$ is sensitive to
the behaviour of $\rho(q)$ around $q=1$. In particular, if the maximum value
of $q$ is $q_M < 1$, one can study the large $w$ behaviour of $P^*(w)$ by taking the 
Laplace transform of equation (\ref{recu}). One finds in that case that $P^*(w)$ decays 
{\it faster} that an exponential:
\be
\log P^*(w) \propto_{w \to \infty} -w^\beta \qquad \beta=\frac{\log N}{\log q_M N} 
\ee
(Notice that $\beta=1$ whenever $q_M=1$, and that $\beta \to \infty$ when $q_M=1/N$:
this last case corresponds to an ordered lattice with no fluctuations). 
In this sense, the exponential tail of $P^*(w)$ is not universal: it requires the 
possibility that one of the $q$ can be arbitrarily close to $1$. This implies that all
other $q$'s originating from that point are close to zero, i.e. that there is a non
zero probability that one grain is entirely bearing on one of its downward neighbours.

The success of the $q$-model with a uniform distribution of the $q$'s is that it
provides a simple explanation for the ubiquitous exponential tail for the distribution
of forces, observed in many experimental and numerical situations. Note in particular
that this exponential tail was observed in a regular packing of grains, suggesting that
very strong heterogeneities in fact exist at the contact level. On the other hand, the 
probability to 
observe very small $w$ is much underestimated by equation (\ref{exptail}): 
see \cite{Huntley,Radjai,fluctuations,pre}. This might be due to the fact that {\it arching}
effects are absent in this scalar model. A generalization of the $q$-model 
allowing for arching was suggested in \cite{CB}, which dynamically generates some sites where
$q_+=1$ and $q_-=0$  (or vice versa). This indeed leads to much higher probability 
density for small weights. For a more detailed discussion of the relation between the
$q$-model the experimental situations, in particular the role of boundaries, see \cite{Jacco2}.

\subsection{The `critical' case}

There is however one special case of particular interest where the results for $P(w)$ 
are qualitatively different. 
Suppose that $q$ is a random variable that only takes the values $0$ or $1$ with probability
$1/2$. This is called the Takayasu model, which is a 
model for directed 
river networks for example: at each site of the lattice a river flowing `south-east' or
`south-west' is randomly
deflected to the left or to 
the right. Rivers coalesce upon meeting. The `source' term $w_0$ here describes a
constant density of `springs' that feed the river network. It can also be seen as a model of 
diffusing and aggregating clusters in a solution with a constant density of `monomers' 
(that again play the
role of the source term). In this case, it turns out \cite{Taka}
that the stationary distribution $P^*(w)$ becomes, for large heights, a power law, 
$P^*(w) \propto w^{-1-\mu}$, 
with $\mu=1/3$ in dimension $d=2$ (i.e. one `spatial' and one `temporal' dimension). 

The exponent $\mu$ was derived analytically but can also understood as follows. Since the
direction of the `rivers' is, at each step, a random variable, the
typical `basin of attraction' of a given site is a parabolic object of height $t$ and 
width $\sqrt{t}$. Therefore, on the order of $t^{3/2}$ `springs' at most can contribute to the 
river flux on a given site; in other words, one expects the distribution $P^*(w)$ to 
take the scaling form:
\be\label{power}
P^*(w) = \frac{1}{w^{1+\mu}}\,  F\left(\frac{w}{t^{3/2}}\right),
\ee
where $F$ is a certain function which falls off fast for large arguments. 
On the other hand, since one must have $\langle w \rangle = w_0 t$ exactly, the exponent $\mu$ 
is fixed:
\be
\int_0^{+\infty} dw\, w\, P^*(w) = w_0 t = t^{3\mu/2} \int_0^{+\infty} du\, \frac{F(u)}{u^{\mu}} \longrightarrow 
\mu=\frac13.
\ee
(Note that the integral over $u$ is convergent for this value of $\mu$.)

Therefore, this model generates a power-law distribution for the local masses, and was proposed
early on as a model of `self-organized criticality'. In fact, the model is critical in the
sense that any deviation of $\rho(q)$ from a sum of two equal delta peaks at $0$ and $1$ 
leads to an exponential truncation of $P^*(w)$ at large $w$'s. Let us add several remarks:

\begin{itemize}

\item One can compute higher moments of the local weight, to find $\langle w^q \rangle
\sim t^{3q/2-1/2}$ for $q > 1/3$. In particular, $\langle w^2 \rangle \sim t^{5/2}$, a
result that we will recover below using direct method. 

\item One can also generalize the model to higher (spatial) dimensions, where one finds $\mu=1/2$ 
for all $d \geq 3$ (with logarithmic corrections in $d=3$). 

\item Consider a rectangular sample of width $W$ and height $H$. What is the order of magnitude of
the largest weight encountered at the bottom ? For a pure power law distribution
such as Eq.~(\ref{power}) with $t \to \infty$, the maximum value of $w$ is known to be of order 
$w_{\max} \sim W^{1/\mu}$. This estimate can obviously only be valid if $w_{\max}$ is found
to be much smaller than the truncation imposed by the function $F$, which is of order $H^{3/2}$. 
This requires $W \ll H^{3\mu/2} \sim H^{1/2}$. However, in this regime where $W$ is smaller than 
the diffusion length, the very argument leading to Eq.~(\ref{power}) breaks down, since the maximum 
weight now scales like $WH$, and not as $H^{3/2}$. Extending the argument, we now find 
that the distribution of 
weights reads:
\be\label{power2}
P^*(w) = \frac{1}{W w} \, G\left(\frac{w}{WH}\right),\qquad (W \ll a H^{1/2}).
\ee
Therefore, (a) for $W \gg H^{1/2}$, the maximum value of $w$ is imposed by the cut-off 
and of order $H^{3/2}$, and not $N^3$. Correspondingly, the participation ratio 
$Y_2=\sum_{i=1} w_i^2/(w_0 HW)^2$ 
that characterizes the 
`localization' of the weight is of order $H^{1/2}/W \ll 1$. In the case (b) $W \ll H^{1/2}$, 
on the other hand, the weight is localized on a finite number of sites, and $Y_2 \sim 1$. 

\end{itemize}

\section{The Scalar Model II: Continuous limit and perturbation theory}
\label{ScalarII}
\subsection{Continuous limit of the scalar model}

Let us focus on the case $N=2$ and define $v$ to be such that $q_\pm (i,j)=(1\pm v(i,j))/2$.
If $v$ is small, the local weight is smoothly varying, and the discrete equation
(\ref{liudiscret}) can then be written in the following differential form:
\be
\label{liudiff}
\partial_t w + \partial_x (vw) = \rho + D_0 \partial_{xx} w\label{eq5}
\ee
where $x=i a$ and $t=j \tau$ are the horizontal and (downwards) vertical variables
corresponding to indices $i$ and $j$, and $a$ and $\tau$ are of the order of the size 
of the grains.
The vertical coordinate has been called $t$ for its
obvious analogy with time in a diffusion problem. $\rho$ is the density
of the material (the gravity $g$ is taken to be equal to $1$),
and $D_0$ a `diffusion' constant, which depends on the geometry of the
lattice on which the discrete model has been defined. This diffusion constant 
is of the order of magnitude of the size of the grains, $a$.

In this model and in the following, we shall assume that the density $\rho$ is not
fluctuating. Density fluctuations could be easily included; it is however easy to
understand that the resulting relative fluctuations of the weight at the bottom of
the pile decrease with the height of the pile $H$ as $H^{-1/2}$, and are thus much 
smaller than those induced by the randomly fluctuating direction of propagation,
encoded by $q$ (or $v$), which remain of order $1$ as $H \to \infty$.

Two interesting quantities to compute are the average `response' $G(x,t|x_0,t_0)$ to a
small density change at point $(x_0,t_0)$, measured at point $(x,t)$, and the correlation
function of the force field, $C(x,t,x',t') = \la w(x,t) w(x',t') \ra_c$, where the averaging
is taken over the realizations of the noise $v(x,t)$.

Equation (\ref{liudiff}) shows that the scalar model of stress propagation is 
identical to that describing tracer diffusion in a (time dependent) flow $v(x,t)$.
This problem has been the subject of many 
recent works in the context of turbulence \cite{Russes}; interesting 
qualitative analogies with that field can be made. In particular, `intermittent' 
bunching of
the tracer field correspond in the present context to patches of large stresses, which may 
induce
anomalous scaling for higher moments of the stress field correlation function.

\vskip 0.5cm
$\bullet$ Fourier transforms.
\vskip 0.5cm

The limit where $a \to 0$ is ill defined and leads to a divergence
of the perturbation theory for large wave-vectors $k$. 
We thus choose to regularize the problem by working within the first Brillouin
zone, i.e., we keep all wave vector components within the interval ${\cal I}=[-\La,+\La]$,
where $\La ={\pi \over a}$. Our Fourier conventions for a given quantity $f$ will
then be the following: \bea
\label{fourier_rules1}
f(x,t) & = & \int_{-\La}^{\La} {dk \over 2\pi} e^{\imath k x} f(k,t) \\
\label{fourier_rules2}
f(k,t) & = & \ell_x \sum_{x=-\infty}^{+\infty} e^{-\imath k x} f(x,t)
\eea
One has to be particularly careful when computing convolution integrals, such 
as $\int {dq \over 2\pi} f_1(q) f_2(k-q)$ which must be understood with limits $-\La+k,
\La$ (resp. $-\La, \La+k$) if $k \geq 0$ (resp. $k \leq 0$). An important example,
which will appear in the response function calculations, is:
\be
\label{convo_q}
\int_{q,k-q \in  {\cal I}}{dq \over 2\pi} q = {\Lambda k \over 2\pi} + {\cal O}(k^2)
\ee
Let us then take the Fourier transform of equation (\ref{liudiff})
along $x$, to obtain:
\be
\label{liuTF}
(\partial_t + D_0k^2) w_k = \rho_k + \imath k \int {dq \over 2\pi} w_q v_{k-q}
\ee
Our aim is to calculate, in the small-$k$ limit, the average response
(or Green) function $G(k,t-t')$ defined as the expectation value of the functional
derivative $\la \de w(k,t)/\de \rho (k,t') \ra$;
and the two points correlation function of $w$, $C(k,t)=\la w(k,t)w(-k,t) \ra$.

\vskip 0.5cm
$\bullet$ The noiseless Green function.
\vskip 0.5cm
The noiseless (bare) Green function (or `propagator') $G_0$ is the solution of the 
equation where
the `velocity' components $v_q$ are identically zero: 
$(\partial_t + D_0k^2) G_0(k,t-t') = \de (t-t')$
which is:
\be
\label{G01}
G_0(k,t-t') = \theta (t-t') e^{-D_0k^2(t-t')}
\ee
In real space, the propagator is simply the heat kernel, 
\be
\label{realG0}
G_0(x,t-t') = {\theta(t-t') \over \sqrt{4\pi D_0(t-t')}}
e^{-{x^2 \over 4D_0(t-t')}}.
\ee
This shows that in the non-disordered scalar model, the stress `diffuses', as already noticed
above in the discrete formulation of the model, see Eq.~(\ref{rw}).

\vskip 0.5cm
$\bullet$ Statistics of the noise $v(x,t)$.
\vskip 0.5cm
The noise term $v$ represents the effect of local heterogeneities in the granular packing. 
The mean value of the noise $v$ is taken to be zero, and its correlation
function is chosen for simplicity  to be of the factorable form
$\la v(x,t)v(x',t') \ra =\sigma^2 g_x(x-x') g_t(t-t')$, where $g_x$ and $g_t$ are noise
correlation functions along $x$ and $t$ axis. We shall take $g_x$ and $g_t$ to be
short-ranged (although this may not be justified: fluctuations in the micro-structure of 
granular
media may turn out to be long-ranged due to e.g. the presence of long stress paths or
arches), with correlation lengths $\ell_x$ and $\ell_t$.
Our aim is to describe the system at a
scale $L$ much larger than both the lattice and the correlation lengths:
$a,\: \tau,\: \ell_x,\: \ell_t \ll L$. This will allow us to
look for solutions in the regime $k, E \to 0$, where $k$ and $E$ are
the conjugate variables for $x$ and $t$ respectively, in Fourier-Laplace space.
However, we shall see below that the limit $a, \tau, \ell_x, \ell_t \to 0$ can
be tricky, and must be treated with care: this is because the noise 
appears in a multiplicative manner in equation (\ref{liudiff}).
For computational purposes, we shall often implicitly assume that the 
probability distribution of $v$ is Gaussian; this might however introduce artifacts 
which we try to discuss.

\vskip 0.5cm
$\bullet$ Ambiguities due to  multiplicative noise. Ito vs. Stratonovitch.
\vskip 0.5cm
In equation (\ref{liuTF}), we have omitted to specify the dependence on the variable $t$.
There is actually an ambiguity in the product term $w_q v_{k-q}$. In the discrete $q$-model
model \cite{Liu}, the $q_\pm$'s emitted from a given site are independent of the value of
the weight on that site. In the continuum limit, this corresponds to choosing $w_q(t)$ to
be independent of $v_{k-q}(t)$, or else that the $v$'s must be thought of as slightly
posterior to the $w$'s (i.e  the product is read as $w_q(t-0) v_{k-q}(t+0)$). In this case, 
the average of
equation (\ref{liuTF}) is trivial and coincides with the noiseless limit; hence  $G=G_0$.
This can be understood directly on the discrete model by noticing that the Green function
$G(i,j|0,0)$ can be expressed as a sum over paths, all starting at site $(0,0)$,
and ending at site $(i,j)$:
\be
\label{Q1}
G(i,j|0,0)=\sum_{\mbox{paths } \cal{P}} \ 
\prod_{\mbox{($k$,$l$)} \in \cal{P}} \: q_\pm(k,l)
\ee
where the $q_\pm(k,l)$ are either $q_+(k,l)$ or $q_-(k,l)$, depending on the  path.
Since each bond $q_\pm(k,l)$ appears only once in the product,
the averaging over $q$ is trivial and leads to:
\be
\label{Q2}
G(i,j|0,0)=\sum_{\mbox{paths } \cal{P}} \ 2^{-j} \equiv G_0(i,j|0,0)
\ee
(Note that this
argument fails for the computation of the correlation function $C$, since
paths can `interfere'. We shall return later to this calculation.)

The above choice corresponds to Ito's prescription in stochastic calculus. 
Another choice (i.e. Stratonovitch's prescription) is however possible,
which corresponds to the proper continuum time limit in the case where the 
correlation length $\ell_t$ is very small, but not smaller than $a$.
In this case, the $w$'s and the $v$'s cannot be taken to be independent.
This is the choice that we shall make in the following.

\subsection{Calculation of the averaged response and correlation functions.}

Two approaches will be presented. The first one, based on
Novikov's theorem, leads to exact differential equations for $G$ and $C$,
which can be fully solved. The second one is a mode-coupling approximation ({\sc{mca}}),
based on a re-summation of perturbation theory. It happens that, for this particular model
where the noise is Gaussian and short range correlated in time, both approaches give
the same results, because perturbation theory is trivial. In other cases, though,
where exact solutions are no longer available, the {\sc{mca}} is in general
very useful to obtain non perturbative results.  

We shall see that the effect of the noise is to widen the diffusion peak:
$D_0$ is renormalized by an additional term proportional to the variance of the noise $v$.

\vskip 0.5cm
$\bullet$ Novikov's theorem. Exact equations for $G$.
\vskip 0.5cm
Novikov's theorem provides the following identity, valid if the $v$ are Gaussian 
random variables:
\be
\label{novi1}
\la w(k,t)v(k',t) \ra =\int_0^t dt' \int dq
\left < {\de w(k,t) \over \de v(q,t')} \right > \la v(q,t')v(k',t) \ra
\ee
Such a term actually appears in equation (\ref{liuTF}), after transformation into an equation
for $G$:
\be
\label{novi2}
(\partial_t+D_0k^2)G(k,t-t')= \de (t-t')
-\imath k {\de \over \de \rho (k,t')} \int {dq \over 2\pi}
\la v(q,t) w(k-q,t) \ra 
\ee
In  the limit where $\ell_x = a \to 0$, the noise correlation is of the form:
$\la v(q,t)v(q',t') \ra =2 \pi \sigma^2 \de (q+q') g_t(t-t')$, with $g_t$
peaked in $t=t'$ such that $f(t')g_t(t-t') \simeq f(t)g_t(t-t')$ for
any function $f$. From formally integrating equation (\ref{liuTF}) between $t'$ and $t$,
one can express the equal-time derivative $\de w / \de v$ as:
\be
\label{novi3}
\left.{\de w(k,t) \over \de v(k',t')}\right|_{t' = t-0}= -\imath k w(k-k',t)
\ee
and thus obtain:
\be
\label{novi4}
(\partial_t+D_0k^2)G(k,t-t') = \de (t-t')
- 2 \pi \sigma^2 k G(k,t-t') \int_0^t dt' g_t(t-t') \int {dq \over 2\pi} (k-q)
\ee
Using the shape of the function $g_t$, the first integral is $1/2$.
The second one is a convolution integral, and its value is
$\Lambda k/2\pi + {\mathcal{O}} (k^2)$ (see equation (\ref{convo_q})). The final
differential equation for $G$ is then, in the small-$k$ limit, 
a diffusion equation with a renormalized diffusion constant:
\be
\label{Dchapeau}
D_R = D_0 + {\sigma^2 \Lambda \over 2}
\ee
It is interesting to note that the model remains well defined in the limit where the
`bare' diffusion constant is zero, since a non zero diffusion constant is induced by
the fluctuating velocity $v$. This would not be true if Eq. (\ref{eq5}) was interpreted 
with the Ito convention,
where the fluctuating velocity would {\it not} lead to any spreading of the average density.

The most important conclusion is thus that, in the present scalar model, stresses propagates
essentially vertically, even in the presence of disorder: taking $\ell \sim a$ (where $\ell_x \sim 
\ell_t \sim \ell$), the response at 
depth $H$ to a small perturbation
is confined within a distance $\propto \sqrt{D_R H}$ from the vertical. Since
$D_R \simeq \ell^2/a$, $\sqrt{D_R H}$ is much less than $H$ in the limit where
$H \gg \ell^2/a$, i.e. when the height of the assembly of grains is much
larger than the grain size. 
 
\vskip 0.5cm
$\bullet$ Exact equations for $C$.
\vskip 0.5cm
Exact equations can also be derived for $C$, following very similar
calculations. From equation (\ref{liuTF}), one can deduce the corresponding
one for $w(k,t)w(-k,t)$. Upon averaging, Novikov's theorem has
to be used on quantities such as $\la w(k,t)v(q,t)w(-k-q,t) \ra$, finally leading to:
\be
\label{C1}
(\partial_t+2D_R k^2) C(k,t) = \hat \sigma^2 k^2 \left[\int {dq \over 2\pi} C(q,t)+\rho^2 t^2\right],
\ee
where $\hat \sigma^2=(2 \pi) \sigma^2$.
Going to Laplace transforms in time leads to:
\be
\label{C2}
(E+2D_R k^2) C(k,E) - C(k,t=0) =  \hat \sigma^2 k^2 \left[\tilde{C}(E)+2\rho^2 E^{-3}\right]
\ee
where $\tilde{C}(E)=\int {dk \over 2\pi} C(k,E) \equiv \int dt\, e^{-Et} C(x=0,t)$. 
This allows one to get a closed equation on $\tilde{C}(E)$:
\be
\tilde{C}(E) = \int {dk \over 2\pi} \frac{C(k,t=0)}{E+2D_R k^2} + 
\int {dk \over 2\pi} \frac{\hat \sigma^2 k^2}{E+2D_R k^2} \left(\tilde{C}(E) + 2\rho^2 E^{-3}\right).
\ee
In the limit where $E \to 0$, the first term on the right hand side is of order $E^{-1/2}$, 
whereas the coefficient of $\tilde{C}(E)$ is equal to $\hat \sigma^2/2D_R$. This shows that one
has to distinguish two cases:
\begin{itemize}

\item $\hat\sigma^2 < 2D_R$. This corresponds, for the discrete model, to the generic case 
where $\rho(q)$ is not the sum of two delta peaks at $q=0$ and $q=1$. The equation for 
$\tilde{C}(E)$ becomes, for $E \to 0$:
\be
\left(1- \frac{\hat \sigma^2}{2 D_R}\right)\tilde{C}(E) \approx \frac{\hat \sigma^2}{2 D_R} 2\rho^2 E^{-3},
\ee
or $\tilde{C}(E) \sim E^{-3}$. Transforming back to real times leads to $C(x=0,t) \sim t^2$, 
which is consistent with the results obtained within the discrete scalar model, where 
the local weight is of order $t$, with relative fluctuations of order one. Once one
knows $C(x=0,t)$, one can obtain, from Eq.~(\ref{C1}), the full function $C(x,t)$. When $x \gg a$,
one finds:
\be
C(x,t) = t^{3/2} {\cal C}\left(\frac{x}{\sqrt{t}}\right), 
\ee
where ${\cal C}(.)$ is a scaling function computed explicitly in \cite{Majumdar}. This results
shows (i) that the correlations extend over distances $\sqrt{t}$, as expected from the 
diffusive nature of the model, and (ii) that $C(x \neq 0,t) \ll C(x=0)$, meaning that 
asymptotically correlations between different sites vanish. The assumption that different 
sites become independent (which is a stronger statement) was already used above to obtain the 
Gamma distribution of the local weights in the scalar model. We see that this assumption is at 
least consistent. 

\item $\hat \sigma^2 = 2D_R$. This corresponds exactly, as shown in \cite{Majumdar} to the critical
Takayasu model where $q$ can only take the values $0$ and $1$ with equal probability. In this
case, a more careful analysis of the limit $E \to 0$ must be performed.  Using the fact that:
\be
\int {dk \over 2\pi} \frac{2D_R k^2}{E+2D_R k^2} = 1 - E \int {dk \over 2\pi} 
\frac{1}{E+2D_R k^2} \sim 1 - \sqrt{E/8D_R},
\ee
we now find that $\tilde{C}(E) \sim E^{-7/2}$, or $C(x=0,t) \sim t^{5/2}$. Again, this is
consistent with the direct scaling analysis presented above. Extending the analysis to $x \neq 0$
now leads to:
\be
C(x \neq 0,t) = t^{2} {\cal C}_c\left(\frac{x}{\sqrt{t}}\right), 
\ee
where the critical scaling function ${\cal C}_c(.)$ can also be computed explicitly 
\cite{Majumdar}, and is different from ${\cal C}(.)$. Note that the asymptotic de-correlation
of different sites still takes place, since $C(x \neq 0,t) \ll C(x = 0,t)$. 

\end{itemize}
\vskip 0.5cm
$\bullet$ Perturbation theory.
\vskip 0.5cm
The above method gives exact results, essentially because $v(x,t)$ is short range
correlated in time: $\delta w/\delta v$ is then only needed at coinciding times, where
it is known, and equal to $1$. This would not be true in general; furthermore Novikov's
theorem requires $v$ to be Gaussian. It is thus interesting to show how a systematic
perturbation scheme can be made to work by the use of diagrams
to represent equation (\ref{liuTF}). The {\sc{mca}} (Mode Coupling Approximation) is
then a particular re-summation scheme of this set of diagrams, which was discussed
in detail in \cite{MCA}, which sometimes provide interesting non perturbative results.

Equation (\ref{liuTF}) is multiplied on the left by the operator $G_0$
(see equation (\ref{G01})), and then re-expressed as follows:
\be
\label{mc1}
w(k,t)=G_0(k,t) \otimes \rho (k,t) -\imath k G_0(k,t) \otimes \int {dq \over 2\pi}
 w(q,t)v(k-q,t)
\ee
$\otimes$ meaning a $t$-convolution product.
This equation can be represented with diagrams as follows: as shown in 
figure \ref{diagdef}, we represent the source $\rho$ by a cross, the `bare'
propagator $G_0$ by a plain line and the noise $v$ by a dashed line. 
\bfig[htb]
\bc
\epsfysize=4cm
\epsfbox{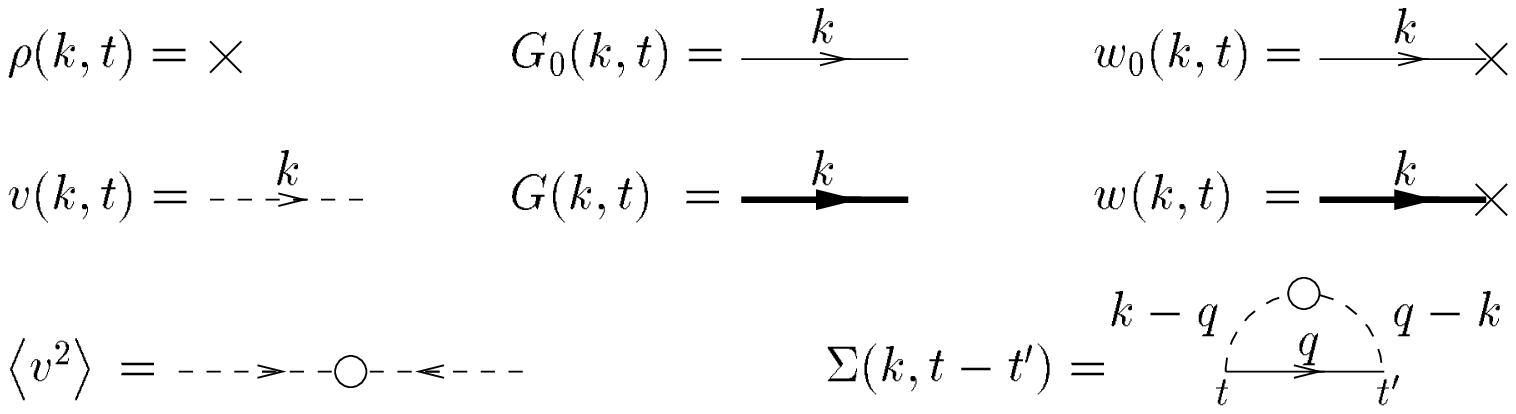}
\caption{\small definition of various diagrams.
\label{diagdef}}
\ec
\efig
The first term of equation (\ref{mc1}), which is the
noiseless solution $w_0$, is then obtained as the juxtaposition of a plain line and
a cross. The arrow flows against time (i.e it is directed from $t$ to $t' < t$). 
The juxtaposition of
two objects means a $t$-convolution product. By definition $w$ is represented by the
juxtaposition of a bold line and a cross (this is consistent with the identification
of a bold line with the full propagator $G$). The diagrammatic version
of equation (\ref{mc1}) is then:
\be
\label{diagequa1}
\includegraphics{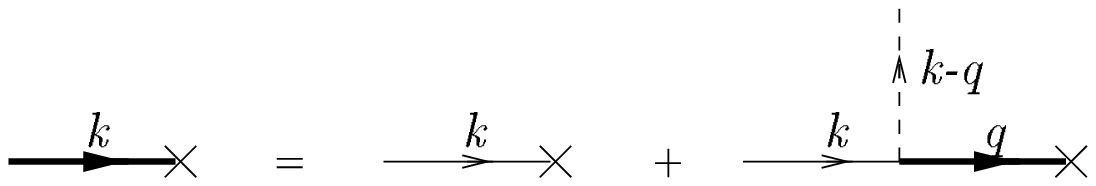}
\ee
The `vertex' stands for $-\imath k \int {dq \over 2\pi}$, the two emerging
wave vectors being $q$ and $k-q$ (node law). One can now iterate this equation.
To second order, one obtains:
\be
\label{diagequa2}
\includegraphics{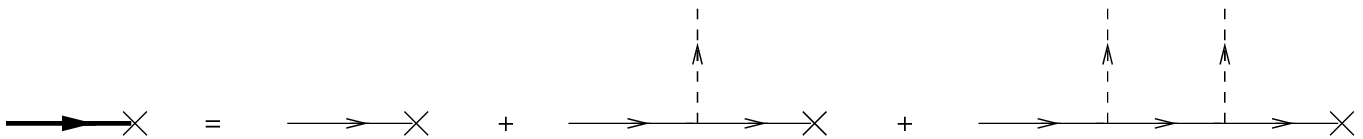}
\ee
The corresponding equation for $G$ is obtained by taking the derivative
$\de / \de \rho$, and averaging over the noise $v$. Since $\la v \ra =0$, the
second diagram vanishes. We represent the noise correlator by
a dashed line with a centered circle (see figure \ref{diagdef}), and 
obtain:
\be
\label{diagequa3}
\includegraphics{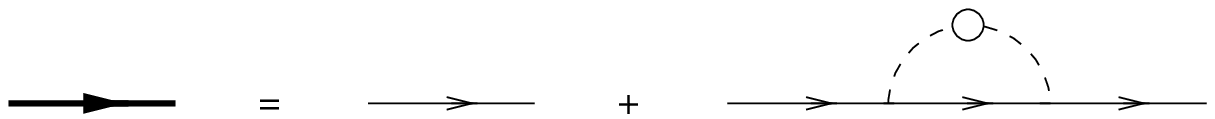}
\ee
or $G = G_0 + G_0 \Sigma G_0$, where $\Sigma$ is called the self-energy
(see figure \ref{diagdef}).  Actually, one can re-sum exactly all the diagrams
corresponding to $G_0 \Sigma G_0$, $G_0 \Sigma G_0 \Sigma G_0$ to obtain 
the Dyson equation $G = G_0 + G_0 \Sigma G$. 

The {\sc{mca}} amounts to replacing the `bare' propagator in the diagram for $\Sigma$ 
by the full
propagator $G$.  (Note that the {\sc mca} is of course exact to second order in perturbation 
theory). We then
obtain a self-consistent equation for $G$:
\be
\label{diagequa4}
\Sigma_{{\sc{mca}}} = G_0^{-1} - G_{{\sc{mca}}}^{-1} = 
\includegraphics{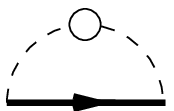}
\ee
Diagrams like the one drawn in figure \ref{diagequa5} are now also
included.
\bfig[htb]
\bc
\epsfysize=2cm
\epsfbox{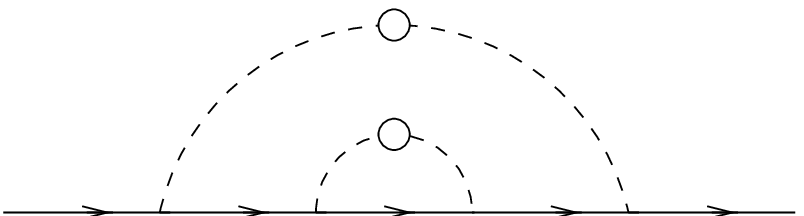}
\caption{\small Example of a diagram included in the {\sc{mca}}. Note that this diagram is zero 
if the noise is $\delta$-correlated in time.
\label{diagequa5}}
\ec
\efig
The self-energy $\Sigma_{{\sc{mca}}}$ can be easily computed, we get
\be
\label{sigma}
\Sigma_{{\sc{mca}}} (k,t-t') = -2 \pi \sigma^2 k \int {dq \over 2\pi} q 
G_{{\sc{mca}}}(q,t-t') g_t(t-t')
\ee
In the special case where $g_t$ is peaked around $t=t'$, we can make the approximation
$G(q,t-t') g_t(t-t') \simeq G(q,0) g_t(t-t') = g_t(t-t')$ (since
by definition $G(q,0)=1$).  We thus get, using equation (\ref{convo_q})),
$\Sigma_{\sc{mca}}(k,t-t')= -\sigma^2 \La k^2 g_t(t-t')$.
The expression for $G_{{\sc{mca}}}^{-1}$ is thus identical to the one obtained
with the exact approach, as can be seen by comparing equation (\ref{novi4}) and
$G_0^{-1}G= 1 + \Sigma G$. 

Note that one can also calculate the influence of a non
zero skewness $\varsigma$, or kurtosis $\kappa$, which are the normalized 
third and fourth cumulant of the noise $v$.
In this case, three and four dashed lines (corresponding to $v$) can be merged, leading to a 
contribution to $D$, of the order of $\varsigma \sigma^3$ or $\kappa \sigma^4$. 

Let us turn now to the calculation of the correlation function
$C(k,t)=\la w(k,t)w(-k,t) \ra$. The basic object which corresponds to 
the self-energy is now the `renormalized source' spectrum $S(k,t,t')$ defined as:
$C=G \otimes S \otimes G$. $S$ is drawn as a filled square.
$S_0$ (empty square) is the correlation function source
term which encodes the initial conditions (see below).
The two first terms of the expansion are
\be
\label{diagequa6} 
\includegraphics{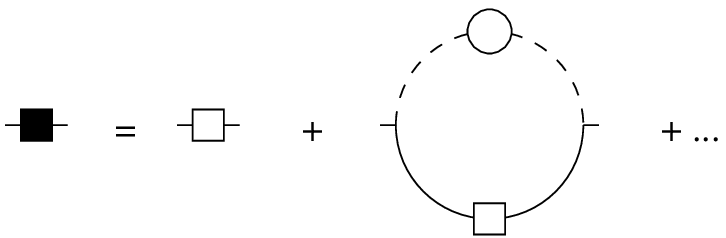}
\ee
Here again, we transform the perturbative expansion into a closed
self-consistent equation for $S$ by replacing $G_0$ and $S_0$ in
(\ref{diagequa6}) by $G$ and $S$ respectively. The final equation for $C$ reads:
\be
\label{diagequa7} 
\includegraphics{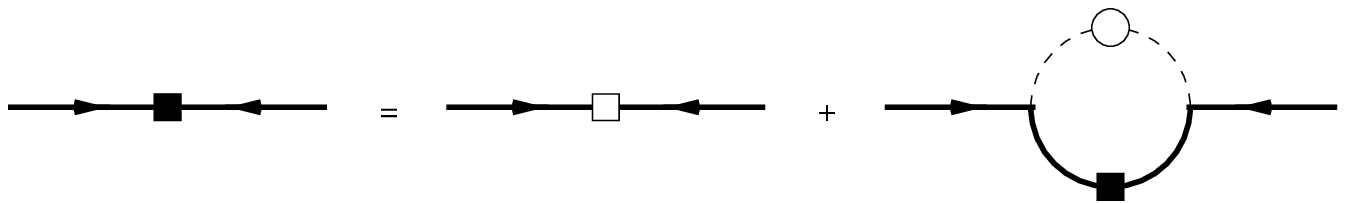}
\ee
or, written explicitly,
\bea
\lefteqn{ C(k,t) = \int_0^t dt' \int_0^t dt''
G(k,t-t')  S_0(k,t',t'')  G(-k,t-t'') + } \nonumber \\
\label{Cmca}
& &  \hat \sigma^2 k^2 \int_0^t dt' \int_0^t dt''
G(k,t-t') \int {dq \over 2\pi} C(q,t',t'') g_t(t'-t'') G(-k,t-t'')
\eea
If we choose the source term to be an overload localized at $t=0$, we get:
$S_0 = \la \rho(k,t') \rho(-k,t'') \ra = C(k,0) \de(t') \de(t'')$.

Using the fact that $g_t$ is peaked around $t'=t''$, we again recover
exactly the equation (\ref{C1}) above, showing again that {\sc mca} is 
exact in this special case.

\subsection{Further results: the un-averaged response function}

The {\it average} Green function described above is thus a Gaussian of zero mean, 
and of width growing as $\sqrt{D_R t}$. However, for a {\it given} environment, 
the Green function is not Gaussian, presenting sample dependent peaks 
(see Figure \ref{repscalar}). Note however that, contrarily to what we shall find
below for the tensorial case, the un-averaged Green function remains everywhere positive.
\bfig[hbt]
\bc
\epsfysize=8cm
\epsfbox{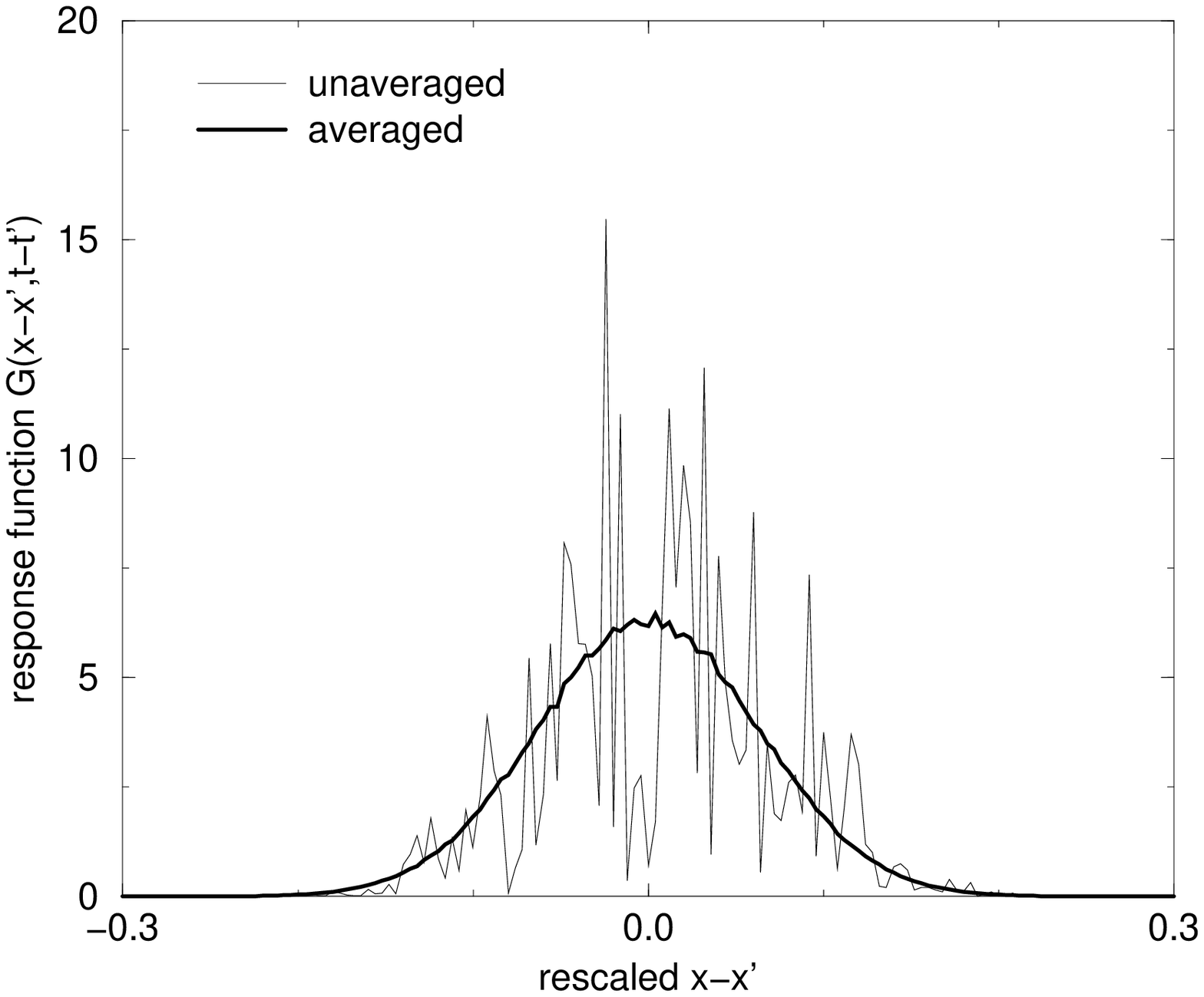}
\caption{\small Averaged (bold line) and un-averaged (thin line) response function of 
the scalar model, obtained numerically by simulating the $q$-model.
One can notice how `non self averaging' is the response function, i.e.
how different it is for a given environment as compared to the average. Note
also that the un-averaged Green function is everywhere positive. \label{repscalar}}
\ec 
\efig 
Furthermore, the quantity $[x](t)$, defined as the displacement of the
centroid of the weight distribution beneath a point source in a given realization:
\be
[x](t) = \int_{-\infty}^{+\infty} dx' \  x' \ \frac{\delta w(x',t)}{\delta \rho(0,0)} 
\ee
typically grows with $t$. More precisely, one can show that:
\be
\la [x](t) \ra =0 \quad {\rm but} \quad \la [x]^2(t) \ra \propto t^{1/2}
\ee
meaning that the `center' of the Green function wanders away from the origin in a sub-diffusive 
fashion,
as $t^{1/4}$. This behaviour has actually been obtained in another context, that of a
quantum particle interacting with a time dependent 
random environment. Physically, the $q$-model can indeed be seen as a collection
of time dependent scatterers, converting in-going waves into outgoing waves with 
certain partition
factors $q_+,q_-=1-q_+$, a problem equivalent to the one dimensional 
Schroedinger equation with a time dependent random potential 
(see the discussion in \cite{DirectedWaves1,DirectedWaves2}). In two dimensions (plus time), the wandering
of the packet center $[x](t)$ is only logarithmic 
(and disappears in higher dimensions \cite{DirectedWaves2}).

Similarly, the participation ratio $Y_2 = \int dx \la w^2(x,t) \ra$ can be computed, and
is found to be $\sim t^{-1/2}$, which means that the
weight is not localized on a finite number of sites in this model when $t \to \infty$. 

\subsection{The scalar model with bias: Edwards' picture of arches}

Up to now, we have considered the mean value of $v$ to be zero, which reflects the
fact that there is no preferred direction for stress propagation. In some
cases however, this may not be true. Consider for example a sandpile built from a point source:
the history of the grains will certainly in-print a certain oriented `texture' to the contact
network, which can be modeled, within the present scalar model, as a non zero value of
$\la v \ra$, the sign of which depends on which side of the pile is chosen. In other words,
the isotropic Edwards assumption for the local stress transmission is expected to break down
when the history of the packing explicitly breaks a symmetry. 

Let us call
$V_0$ the average value of $v$ on the $x\ge 0$ side of the pile, and $-V_0$ on the other
side. The differential equation describing propagation now reads, in the absence of disorder:
\be
\label{epe}
\partial_t w + \partial_x \left [V_0 \ \mbox{sign}(x)w \right] = \rho + 
D_0 \partial_{xx} w
\ee
For a constant density $\rho=\rho_0$, and for $D_0=0$,
the weight distribution is then the following: 
\bea
\label{eps}
w(x,t) = & {\rho_0x \over V_0}        & \qquad   \mbox{ for  $0 \le x \le V_0t $} \nonumber\\
w(x,t) = & {\rho_0(ct-x) \over c-V_0} & \qquad  \mbox{ for  $V_0 t\le x \le ct $}
\eea
where $c=1/\tan \phi$ ($\phi$ is the angle made by the slope of the
pile with the horizontal $x$ axis). For $D_0 \neq 0$, the above solution is smoothed.
In any case, the local weight reaches a {\it minimum} around $x=0$.  Equation (\ref{epe}) gives a
precise mathematical content to Edwards' model of arching in sand-piles \cite{Edwards},
as the physical mechanism leading to a `dip' in the pressure distribution \cite{Smid}.
As discussed elsewhere \cite{WCCB,FPA}, this can be taken much further within a
tensorial framework (see Section \ref{tensorial}).

Equation (\ref{epe}) with noise can in fact be obtained naturally within an extended $q$-model,
with an extra rule accounting for the fact that a grain can slide and lose contact with 
one of its two downward neighbour when the shear stress is too large \cite{CB}. This generically 
leads to arching; in the sandpile
geometry and for above a certain probability of (local) sliding, the effective `velocity' $V_0$
becomes non zero and the weight profile (\ref{eps}) is recovered \cite{CB}. However, this extra
sliding rule implicitly refers to the existence of shear stresses, which are absent in the scalar
model, but which are crucial to obtain symmetry breaking effects modeled by a non zero $V_0$ (see 
also the discussion in section~\ref{tensorial}.
It is thus important to consider from the start the fact that stress has a tensorial,
rather than scalar, nature. This is what we investigate in the following sections.

\section{Static indeterminacy; elasticity and isostaticity}

\subsection{Elasticity and response functions}

As mentioned in the introduction, the sole equilibrium equations are not sufficient to 
determine the stress tensor of an arbitrary material. In $d=2$, one has two equations and
three independent components of the stress tensor. In $d=3$, there are three equations for
six independent components of $\sij$. In elastic materials, this indeterminacy is lifted when 
one adds the constraint that the stress tensor is linearly related to the strain. 
The most general linear relation between stresses and strains is given by:
\be
\label{s.u}
\sigma_{ij}=\lambda_{ijkl}u_{kl}
\ee
where $\sigma_{ij}$ denotes the components of the stress tensor, $u_i$ is the displacement field, 
$u_{ij}=\frac{1}{2}(\partial_j u_i+\partial_i u_j)$ those of the strain
tensor, and summation over repeated indices is implied. The four index tensor $\lambda_{ijkl}$
satisfies certain symmetry conditions \cite{LL}.

In order to close the problem for the stress tensor, one imposes a condition of `compatibility',
which in $d=2$ reads:
\be
\label{compa}
\partial_z^2 u_{xx}+\partial_x^2 u_{zz}-2\partial_x\partial_z u_{xz}=0,
\ee
resulting  simply from the fact that the tensor $u_{ij}$ is built with the derivatives of 
a vector $u_i$. This is enough to find a closed equation for the stresses (in $d=2$) \cite{Otto}: 
\be
\label{sijequation}
\left ( \dr^4_z + t\dr^4_x + 2 r \dr^2_x\dr^2_z \right ) \sij =  0
\ee
where the two independent coefficients $t$ and $r$ can be expressed in terms of the 
components of $\lambda_{ijkl}$. {\it Isotropic} elasticity corresponds to $r=t=1$. 

Expanding the stresses in Fourier modes, it is easy to see that the
solutions of the equations (\ref{sijequation}) are of the
form
\bea
\szz & = &          \iqt \sum_{k} a_k(q) \, e^{iqx + iX_kqz},
\label{szzFourier} \\
\sxz & = & C_{xz} - \iqt \sum_{k} a_k(q) \, X_k   \, e^{iqx + iX_kqz},
\label{sxzFourier} \\
\sxx & = & C_{xx} + \iqt \sum_{k} a_k(q) \, X_k^2 \, e^{iqx + iX_kqz},
\label{sxxFourier}
\eea
where $C_{xx}$ and $C_{xz}$ are constants. From equation (\ref{sijequation})
we see that the $X_k$ are the roots of the following quartic equation
\be
X^4 + 2 r X^2 + t = 0,
\label{equaX}
\ee
which has four solutions:
\be \label{rootX}
X = \pm \sqrt{- r \pm \sqrt{r^2 - t}}.
\ee
Hence the index $k$ runs from $1$ to $4$.  The four functions $a_k(q)$
and the constants $C_{xx}$ and $C_{xz}$ must be determined by the boundary
conditions. A particularly interesting boundary condition is when one 
imposes a localized force at the top surface of the material. The shape 
of the stress {\it response function} to such a localized force will be of
central importance in the following discussion. One can establish the existence of various 
`phases' in the $r,t$ plane in terms of the
shape of the response function, as obtained from the calculations 
presented in \cite{Otto}. In that plane, the line $t=r^2$, for $r<0$,
separates the so called {\it hyperbolic} and the {\it elliptic} regions. For $t > r^2$, 
the above roots $X_k$ are complex whereas for $t < r^2$ and $r >0$ the roots $X_k$ are 
purely imaginary. These two regions correspond to the {\it elliptic} regime, which is
in fact the only accessible one in the context of classical elasticity where the 
coefficients $r$ and $t$ are constrained by the fact that the undeformed state is
a minimum of the elastic energy.  As shown in details in \cite{Otto}, the response function 
has a unique, broad peak of width growing linearly with depth in the region $t < r^2$ and $r >0$,
whereas the response function becomes {\it double peaked} in the region 
$r<0$, $t > r^2$ (with a width again scaling linearly with depth). As one
approaches the line $t=r^2$, the two peaks become narrower and
narrower before becoming two delta-function peaks exactly on the
transition line.  At this point and below the transition, the system is {\it hyperbolic}; 
this limit behaviour will actually emerge naturally below in the context of granular
materials.

\subsection{Indeterminacy at the grain level and  isostaticity}

Elasticity theory can also be seen as the long-wavelength description of a network 
of beads and springs, for which the local equilibrium equations are fully determined
at each node. When the network is disordered, the theoretical difficulty is to compute 
the effective elastic constants in terms of the probability distribution and 
correlations of the microscopic springs. The same problem arises when one wants to
compute the effective conductivity, or the effective permittivity, etc. of a composite 
material. But in all these problems, the {\it microscopic} equations are sufficient to
solve the problem in principle. 

For an assembly of grains, this is not the case. The indeterminacy of the static equilibrium
exists already at the grain level (see the simple case discussed in section~\ref{halsey}). 
In principle, one
should describe in details the microscopic history of each contact in order to determine
the precise configuration of forces within a given packing. There are however special cases
where this is not the case, and where all contact forces are fully determined by the packing 
geometry. These situations are called isostatic, and play a special role. These equilibria 
are in some sense critical since the opening of one contact necessarily leads to some 
rearrangements. Some arguments have been put forward to suggest that an assembly of grains 
relaxing towards static equilibrium will most probably stop {\it as soon as they are 
stable}, i.e., in one of these isostatic states \cite{blumenfeld}. 
A similar statement is actually {\it exact} in the
context of mean-field spin-glasses, where the equilibrium states reached dynamically are
marginally stable, in the sense that the spectrum of the eigenvalues $\lambda$ of the Hessian 
(matrix of 
second derivatives of the energy) vanishes precisely at $\lambda=0$. Before discussing the
validity of the idea that these marginal states play a special role in granular media, let
us first discuss the geometrical conditions necessary for isostaticity \cite{Halsey2}.  

Consider {\it frictionless} grains in two dimensions. There are, per grain, two equilibrium
equations since the torque is automatically zero, and one (normal) force per contact that
must be determined. If $N$ is the total number of grains and $N_c$ the total number 
of contacts, the number of unknowns is $N_c$ and the number of equations in $2N$. Therefore, one 
can (generically) find static solutions only if $N_c \geq 2N$. Since each contact concerns two
grains, the average number of contact per grain is $n_c=2N_c/N$ and the condition for the existence of 
solutions is $n_c \geq 4$. The marginal case is when $n_c=4$, where the number of unknowns is
equal to the number of equations, and corresponds to the isostatic case. The hyper-static case corresponds to 
a strict inequality. If the friction coefficient is non zero, 
then the zero-torque condition provides a third equation for each grain. If we call $\varphi$ the 
fraction of contacts where friction is fully mobilized (i.e. such that $|F_T|=\mu F_N$), one has 
$\varphi \cdot N_c + (1-\varphi) \cdot 2N_c$ unknowns (one per mobilized contact, two for 
un-mobilized contacts). The stability condition now reads  $n_c(1-\varphi/2) \geq 3$. (Note that
the frictionless case corresponds to $\varphi=1$). In three dimensions, the same argument leads to
$(3-\varphi)n_c \geq 12$, the isostatic case corresponding to an equality. 
The corresponding `stability' diagram is shown in Fig. \ref{nc}. 

\bfig[hbt]
\bc
\epsfysize=8cm
\epsfbox{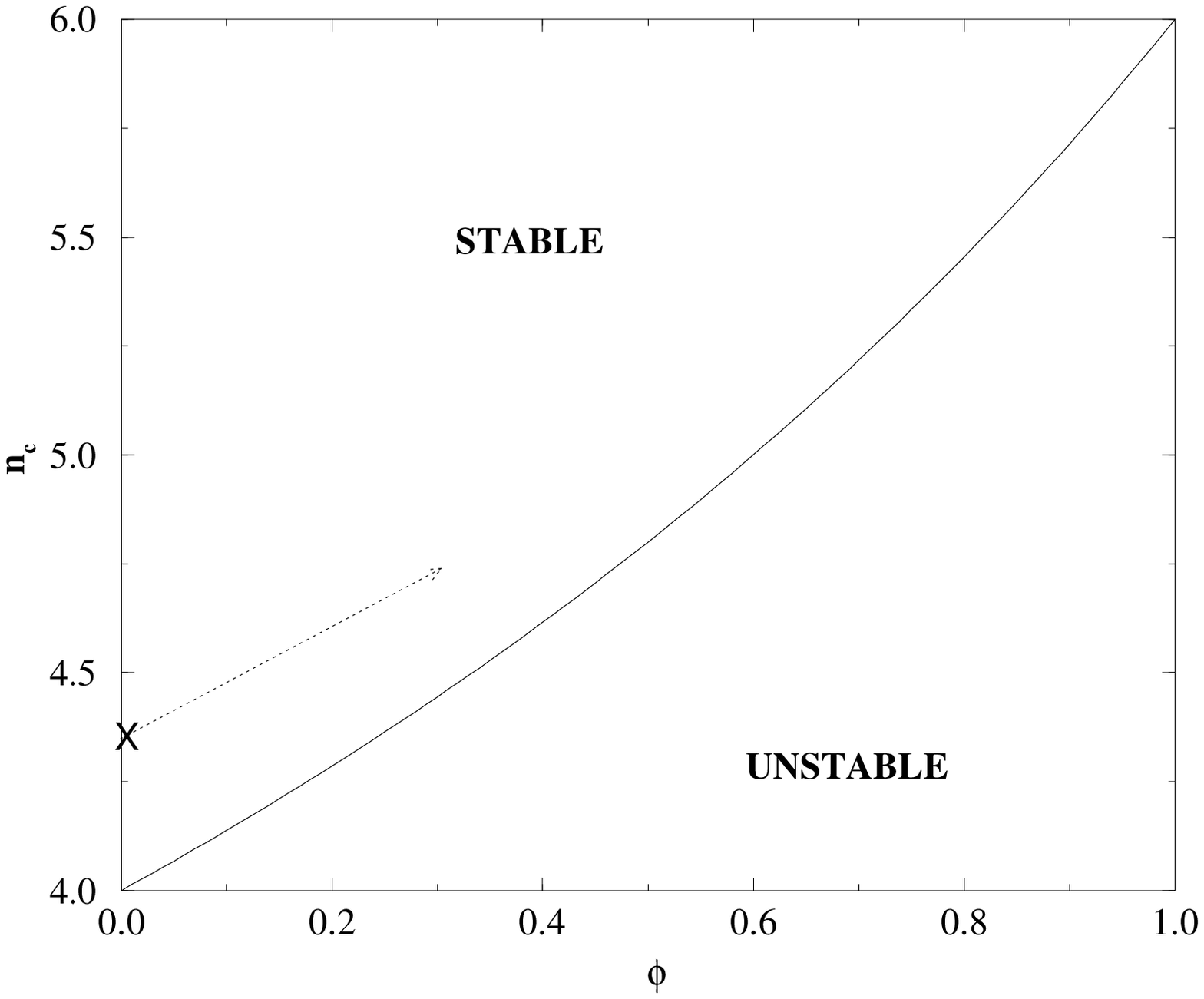}
\caption{\small Stability diagram in the plane $\varphi,n_c$ for three 
dimensional assemblies. The plain line is the isostatic line that separates
stable packings from unstable packings. The point $\varphi=1$, $n_c=6$ corresponds
to frictionless particles. The cross for $\varphi=0$ represents the numerical
result of \protect\cite{Halsey2}; the dotted arrow is a possible path of the packing
when vibrated.}
\label{nc}
\ec \efig 

\subsection{Numerical simulations and Edwards' assumption}

So, are three dimensional packings of grains, obtained by letting the grains lose their 
kinetic energy
and come to rest, generically isostatic? The only quantitative study we are aware of is 
that of Silbert et 
al. \cite{Halsey2}, where these authors perform Molecular Dynamics simulation of 
{\it monodisperse} 
grains with a specific form of contact 
dynamics and a certain energy dissipation coefficient at each collision. The results 
reported in \cite{Halsey2}
are compatible with isostaticity for frictionless spheres, $\mu=0$. In this case, the 
equilibrium packings are indeed
found to obey the condition $n_c = 6$, as expected from the general results 
of \cite{Roux,Moukarzel}. However, when the friction coefficient is non zero, 
these authors find 
that the static configurations are such that (a) the fraction of fully mobilized contacts 
is $\varphi=0$ and 
(b) the number of contacts per grain seems to saturate, in the limit of hard grains, to a 
value of $n_c  > 4$,
suggesting that the packing is not isostatic. The value of $n_c$ appears to 
depend significantly on the value of the friction coefficient and the
restitution coefficient; it appears from their data that smaller restitution
coefficients (i.e. more damping) lowers the value of $n_c$, perhaps down to the isostatic 
limit for large damping. This dependence on
the details of the dynamics indicates that
no universal statement about the statistics of `native' packings 
(i.e. obtained without further tapping) can be made. Again, the simplistic situation discussed in section~\ref{halsey} would provide a useful benchmark.

What happens when one of these native states (possibly hyperstatic) is vibrated? 
Does the packing wander inside the
stable regime (see Fig.~\ref{nc}) or remains near the isostatic boundary ? Here again, 
it is useful to recall the 
results of mean field p-spin glasses, where the `vibrations' (temperature) keep the 
system along the ridge 
of marginally stable states. In this case, the reason is the exponential dominance of 
the number of these states
over the `deeper' (more stable) ones. Therefore, even if blocked states are a priori 
equiprobable (as postulated
by Edwards), the most probable situation is to observe the system in a marginal state. 
If this argument can be 
transposed to granular packings, then ideas that Edwards expressed in different contexts 
(i.e; that blocked states
are equiprobable and that only marginal (isostatic) states are important \cite{blumenfeld}) 
would be reconciled. 
It would be
very interesting to compute the number of metastable states as a function of the isostatic 
index $n_c$ in some
(possibly artificial) model (on this point, see the attempt in \cite{Monasson}).

\section{A stress-only approach to granular media}

We now turn to the discussion of some plausible `stress-only' closure schemes for the static 
equilibrium of granular materials. We first start by a natural generalization of the scalar 
$q$-model
to account for the vectorial nature of the forces. Then we show how the results of this 
`vectorial'
$q$-model can be interpreted more generally in terms of symmetry arguments.

\subsection{A vectorial $q$-model}
\label{tensorial}

It is useful to start with a simple toy model for stress propagation, which
is the analogue
of the scalar model presented in section~\ref{ScalarI}. We now consider the case of three
downward neighbours (see figure \ref{threeleg}), for a reason which will become
clear below. \bfig[hbt]
\bc
\epsfysize=4cm
\epsfbox{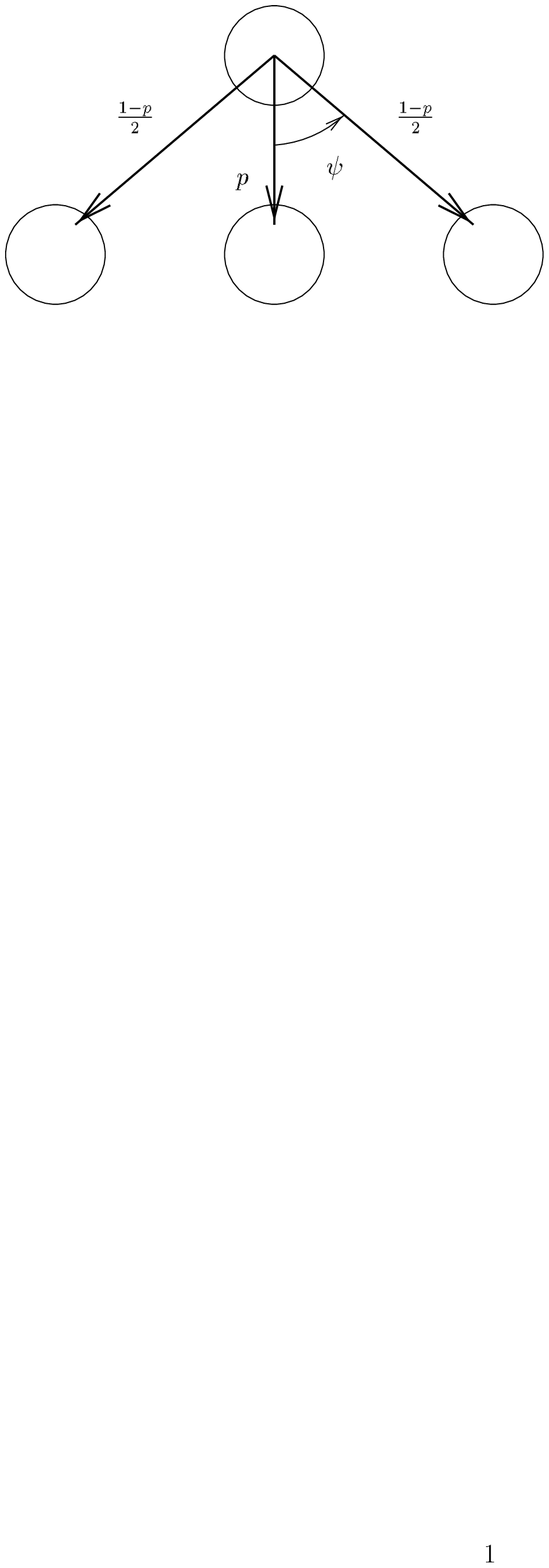}
\caption{Three neighbour configuration. Each grain transmits two force components
to its downward neighbours. A fraction $p$ of the vertical component is
transmitted through the middle leg.} \label{threeleg}
\ec
\efig
Each grain transmits to its downward neighbours not one, but two force
components:
one along
the vertical axis $t$ and one along $x$, which we call respectively $F_t(i,j)$
and $F_x(i,j)$. For simplicity, we assume that each `leg'
emerging from a given grain can only transport
the force parallel to itself. This assumes that each contact is frictionless.
More general transfer rules can be considered, where the forces are chosen, {\it \`a la}
Edwards, 
randomly within the space of solutions -- see e.g. \cite{Eloy,Socolar,Narayan2,Zucker}.

We assume that a fraction $p$ of the vertical force travels through the middle leg.
Correspondingly, a fraction $q_+=(1-p)(1+\varepsilon)/2$ travels through the right leg,
and a fraction $q_-=(1-p)(1-\varepsilon)/2$ through the left leg, so that $p+q_++q_-=1$. 
(Some anisotropic 
generalizations will be discussed below). Since the total force
in these legs is oriented in the direction $\psi$, where $\psi$ is the angle between grains, 
defined in figure \ref{threeleg}, 
the balance of the horizontal force
on the grain $i,j$ imposes a specific value for $\varepsilon$:
\be
\varepsilon \equiv \frac{1}{(1-p) \tan \psi} \frac{F_x(i,j)}{F_z(i,j)}.
\ee
We therefore discover a very important consequence of the existence of a shear stress $F_x$,
which was totally absent from the scalar $q$-model or had to be put by hand: 
any non zero value of $F_x$ necessarily 
leads to $q_+ \neq q_-$ and biases the propagation of $F_z$ in the direction of $F_x$. 
This 
coupling between the two components of the force is at the origin of all the interesting
physics discussed below. 

Using the value of $\varepsilon$ to compute the propagation of the forces from one layer
to the next, we find:
\bea
F_x(i,j) & = & \frac{1}{2} \left[F_x(i-1,j-1)+F_x(i+1,j-1)\right] \nonumber \\
         &   & + \frac{1}{2} (1-p) \tan \psi
\left[F_t(i-1,j-1)-F_t(i+1,j-1)\right] \label{wemicro1}\\
F_t(i,j) & = & w_0 + p F_t(i,j-1) + \frac{1}{2}(1-p)
\left[F_t(i-1,j-1)+F_t(i+1,j-1)\right] \nonumber \\
         &   & +  \frac{1}{2 \tan \psi}
\left[F_x(i-1,j-1)-F_x(i+1,j-1)\right] \label{wemicro2}
\eea
Taking the continuum limit of the above equations leads to:
\bea
\label{equiF1}
\partial_t F_t + \partial_x F_x & = & \rho + a \frac{1-p}{2}\, \partial^2_{tt} F_t \\
\label{equiF2}
\partial_t F_x + \partial_x \left [ c_0^2 F_t \right ] & = & \frac{a}{2}\, \partial^2_{xx} F_x
\eea
where $c_0^2 \equiv (1-p) \tan^2 \psi$, $a$ is the size of the grains, and 
we have kept the second order diffusion terms, which would be the only remaining 
term in the isotropic scalar description, and plays the role of a smoothing term for the 
(singular) solutions found below. Comparing Eq.(\ref{equiF1}) with Eq.~(\ref{epe}) 
(with $w = F_t$),
we see that the `velocity' term introduced by hand in the Edwards model is indeed a 
consequence of the 
local shear. 

Eliminating (say) $F_x$ between the
above two equations leads to a {\it wave equation} for $F_t$ (up to a
diffusion term which becomes small in the large scale limit), where the vertical coordinate $t$ 
plays the r\^ole of time and $c_0$ is the equivalent of the `speed of light'. In particular, 
the
stress does not propagate vertically, as it does in the scalar model, but rather
along two rays, each at a {\it non zero angle} $\pm\varphi$ such that
$c_0=\tan\varphi$. Note that $\varphi
\neq \psi$ in general (unless $p=0$); the angle at which stress propagates has
nothing to do with  the underlying lattice structure and can take any value
depending on the local rules for force transmission. The three-leg model was chosen 
to illustrate this particular point: the number of legs is irrelevant in the large scale
limit, and the wave structure of the resulting equation is universal. 

\subsection{A constitutive relation between stress components} 

The above derivation can be reformulated in terms of classical continuum
mechanics
as follows. Considering all stress tensor components $\sigma_{ij}$, the
equilibrium equation reads,
\bea
\label{equi1}
\partial_t \szz + \partial_x \sxz & = & \rho \\
\label{equi2}
\partial_t \szx + \partial_x \sxx & = & 0
\eea
Identifying the local average of $F_t$ with $\szz$ and that of $F_x$ with
$\szx$,
we see that the above equations (\ref{equiF1}, \ref{equiF2}) are actually
identical to
(\ref{equi1}, \ref{equi2}) provided $\szx=\sxz$ (which corresponds to the
absence of local torque) and, more importantly, that there exists a relation
between the vertical and horizontal components of the stress tensor:
\be
\sxx = c_0^2\, \szz \label{bcc}
\ee
This relation between normal stresses was postulated
in \cite{BCC} as the simplest ``constitutive relation" among stress components,
obeying the correct symmetries, that one can possibly assume.
The term ``constitutive relation" normally refers to a relation between
stress and strain, but the model under discussion has no strain variables
defined. This particular choice can be interpreted as a {\it local}
Janssen approximation \cite{Janssen}, in analogy with the assumption made by Janssen in
1895 in order to describe stresses in silos, where the {\it average} vertical stress 
at a given altitude is postulated to be proportional to the average horizontal stress.
We return later to a more detailed discussion of this type of
closure equations.
In the present case, the parameter $c_0^2$ encodes relevant details of
the local
geometry of the packing (friction, shape of grains, etc.) and may thereby
depend on
the {\it construction history} of the grain assembly. Only for simple,
`homogeneous'
histories (such as constructing a uniform sand-bed using a sieve) will
$c_0^2$ be
everywhere constant on the mesoscopic scale. Even then, unless an ordered
packing is
somehow created, local fluctuations of $c_0^2$ will always be present. The 
influence of these fluctuations will be analyzed below.

\subsection{Some simple situations}

Using Eq.(\ref{bcc}) in Eqs. (\ref{equi1}, \ref{equi2}) we find that these can be rewritten as:
\bea
\label{equi3}
\partial_u ( \sxz - c_0 \szz) & = & -\frac{\rho}{2} \\
\label{equi4}
\partial_v ( \sxz + c_0 \sxz) & = & \frac{\rho}{2},
\eea
where $u = x- c_0 t$ and $v= x +c_0 t$. This shows that in this framework, the forces are 
transported along the characteristics, which is what the word `force chain' usually implies 
\cite{EandO,prl}.
More precisely, the solution of Eqs. (\ref{equi1}, \ref{equi2}) is obtained by projecting 
any `initial' force (i.e. present at altitude $t=0$) onto the two characteristics, and 
propagating and augmenting each component by an amount $(\rho/2){\rm d} t$, independently along 
these characteristics.

The simplest situation is that of an infinitely wide layer of sand, of depth
$H$, with a localized ($\delta$-function) overload at the top. The additional weight at the
bottom then defines the {\it response function} of the wave equation,
which, in two
dimensions, is the sum of two $\delta$ peaks localized at $x=\pm c_0 H$. These $\delta$ peaks
are actually diffusively broadened by the second order terms present in Eqs.~(\ref{equiF1}, 
\ref{equiF2}). This two-peak response function is notably 
different from  the response function of an isotropic elastic body, for which the response
function is a single hump of width proportional to the height $H$. However, for anisotropic
elasticity, a two peak response function {\it can} be observed \cite{Goldhirsch,Rava,Otto}, but 
both peaks have a width scaling linearly with $H$, and not as $H^{1/2}$ for an hyperbolic
equation. The question of the response function is therefore of crucial importance, and 
will be discussed in details below. Recent attempts to determine the response 
function experimentally seem to favor an elastic like response, at least for the strongly 
disordered systems (see section~\ref{response}). 

Next, one
can consider the sandpile geometry. For a pile at repose, the position of
the free
surfaces are
$x =\pm c z$, where $c=\cot\phi$ with $\phi$ the repose angle. On these
surfaces,
all the stresses vanish. This boundary
condition is then (for given $c_0$ and $c$) sufficient to solve for the
stress field
everywhere in the pile.  One then finds that the
vertical normal component of the stress is piecewise linear as a function
of $x$. In
particular, for $-c_0 H
\leq x \leq c_0 H$, $\szz$ is {\it constant}. Therefore, in two dimensions, this
model \cite{BCC} predicts a flat-topped stress profile rather than a hydrodynamic 
pressure hump or a dip. Such a flat top is in fact observed when building a pile from a
uniform rain of grains.

For a pile created by depositing grains from above (for example by sieving
sand onto
a disc) it is natural to expect the free surface to be a slip plane. (This
is a plane
across which the stress components saturate the Mohr-Coulomb condition -- see below.)
Interestingly, this provides a relation between $c_0$ and the friction angle
$\phi$, which reads: $c_0^2=1/(1+2\tan^2\phi)$ (note that since $c=1/\tan \phi$,
one has automatically $c > c_0$). Under these conditions one finds that the `plastic'
region (where the Mohr-Coulomb condition is saturated) extends (in two dimensions) inward from the
surface to encompass the outer `wings' of the pile (i.e. $c_0 z \leq |x| \leq
cz$). This follows from the solution of the model
and is not
an {\it a priori} assumption, of the kind commonly made in elastoplastic
modeling
(e.g., \cite{Cantelaube,Savage}; see also the discussion in \cite{Narayan}). 
In three dimensions, a second closure relation is required
\cite{BCC}, but in all cases the stress profile has a broad maximum at the
center of
the pile. Now, however, the Mohr-Coulomb condition is only saturated in the
immediate vicinity of the free surface -- the `plastic' region has zero volume
in three dimensions \cite{BCC,FPA}.

\subsection{Symmetries and Constitutive Relations}

Although above it was motivated in the context of a specific microscopic model of force 
transfer,
the linear constitutive relation (\ref{bcc}) can be
viewed, independently of any microscopic model, as the
simplest closure equation compatible with the symmetries of the problem.
The latter
include a {\it local} reflection symmetry in which $x-x_0$ is changed to 
$x_0-x$
(with $x_0$ an arbitrary reflection plane) and also a form of ``dilational"
symmetry on the stress
known as RSF (``radial stress field") scaling. RSF scaling depends on the
absence of
any characteristic stress scale, which follows if the Young's modulus of
the grains
is sufficiently much larger than any stresses arising in the granular
assembly being
studied. Such scaling, which requires the stress distributions beneath piles of
different heights to have the same shape, is quite well confirmed in some
(but not
all) experiments on conical sand-piles
\cite{Smid,Huntley,Savage}.

Even with these two symmetries, one
can consider more complicated (nonlinear) constitutive relations among stresses,
which must be of the form
\cite{BCC}:
\be
\sxx = c_0^2 \szz \,{\cal F}\left(\frac{\sxz^2}{\szz^2}\right) \label{closgen}.
\ee
Note that in general, such a constitutive relation violates rotational symmetry, 
and can only describe an anisotropic pile (for example, built from a rain of grains under 
gravity).
The only rotationally invariant stress only closure scheme can only involve the two
invariants of the stress tensor, namely ${\rm Tr}\,\sigma$ and ${\rm Det}\,\sigma$. From 
dimensional analysis, this relation can only be of the form:
\be\label{MC}
\cos^2 \phi ({\rm Tr}\,\sigma)^2 =  4  {\rm Det}\,\sigma,
\ee
where the specific choice of the coefficient is such that we recognize the usual
Mohr-Coulomb relation.  Eq. (\ref{MC}) indeed means that there exists a choice of
orthogonal axis $m,n$ such that:
\be
\left|\frac{\sns}{\snn}\right| = \tan \phi =\mu,
\ee
where $\mu$ is the internal friction coefficient. One can easily show that this relation is of the 
general form Eq.~(\ref{closgen}), for a particular form of ${\cal F}$, since Eq.~(\ref{MC}) 
can also
be rewritten such as:
\be
c_0^2 \, {\cal F}(u) = {1\over \cos^2\phi}\left[
\sin^2\phi+1 \pm
2\sin\phi\,\,\sqrt{1-\cot^2\phi\;\;u}\right]\label{ife}
\ee

Viewed as a
constitutive equation, the 
Mohr-Coulomb relation defines a rigid-plastic model whose physical
content is
to assume that, everywhere in the material, a plane can be found across
which slip
failure is about to occur, hence the name ``incipient failure everywhere",
({\sc ife}), given to this model \cite{BCC,WCCB,FPA}.

All closures of the form (\ref{closgen}) lead to  hyperbolic equations for
stresses, although in the general case the characteristic directions of
propagation
(the `light rays' of the corresponding wave equation) depend on the loading and
therefore vary with position.

An interesting situation arises when local reflection symmetry is broken.
This is
the case, for example, in sand-piles created by pouring from a point source onto
a rough surface -- which is the usual mode of construction. In such a pile, all
grains arriving at the apex of the pile roll (in two dimensions) either to the
right or to the left. The two halves of the pile therefore have different
construction histories that are mirror images of each other. This violates local
reflection symmetry, and in general permits constitutive
equations such as:
\be
\sxx = c_0^2 \szz\, {\cal G}\left(\frac{{\rm{sign}}(x)\sxz}{\szz}\right)
\ee
The simplest case (found e.g. by expanding ${\cal G}$ to first order in the
shear to
normal stress ratio) corresponds to a family of (quasi-) linear constitutive
relations \cite{FPA}:
\be
\sxx = c_0^2 \szz + \nu\, {\rm{sign}}(x)\sxz \label{osl}
\ee
The previous, symmetrical, case has $\nu =0$. For nonzero $\nu$,
(\ref{osl}) again
leads to a wave equation, although this time {\it anisotropic}, in the
sense that
the two characteristic rays make asymmetric angles to the vertical axis. 
 Note also that
$x=0$ is a singular line across which the directions of propagation change
discontinuously.

Microscopically,
$\nu \neq 0$ leads to an unequal sharing of the weight of a grain
between the two characteristic rays propagating downward from it.
Such a model can indeed be obtained from rules such as those in figure
\ref{threeleg} simply by having an asymmetric rules for partitioning the forces between
the supporting grains, for example choosing two different angles $\psi_+$ and $\psi_-$.
The symmetric case corresponds to $\nu=0$ and $\nu \propto \tan \psi_+ -
\tan \psi_-$. For $\nu < 0$, most of the weight travel {\it outwards} on the leg with the
smallest opening angle; this provides, within a fully
tensorial model, a mathematical description of the tendency to form arches, as developed by
Edwards for the scalar case.

Solving these anisotropic wave equations for sand-piles in
two dimensions one again finds for
$\szz$ a piecewise linear function, which now has a maximum at $x=0$
when $\nu
> 0$, but a minimum for
$\nu < 0$, in accord with the arching picture mentioned above. If one
furthermore imposes, as above, that the free surfaces are slip planes, one finds
a relation between $c_0^2$, $\nu$ and $\phi$. One again
finds the result that the material throughout the outer wings of the pile
(exterior
to the triangle formed by the characteristics passing through the apex) are at
incipient (Mohr-Coulomb) failure.

\subsection{Boundary conditions and `fragility'}

All the above closure schemes lead to hyperbolic equations, which crucially differ from the elliptic equations encountered in elasticity theory when
{\it boundary conditions} are considered. Hyperbolic equations indeed require only  
`half' of the boundary conditions to be specified, the other half being 
{\it determined} by `propagating' these known boundary conditions along the
characterisitcs through the
system. On the contrary, an elliptic equation (such as Laplace's equation) 
requires the stress (or
the displacement) on all the surfaces of the body to be specified. If one
insists on applying all the boundary conditions appropriate to an
elastic body,
then in general no solution will exist for the hyperbolic equations that
correspond to a particular choice of constitutive relation. If these 
boundary conditions are `incompatible' in this sense, then within an hyperbolic 
model, the material ceases to be in static
equilibrium.
This is not different from the corresponding statement for a fluid;
if boundary conditions are applied that violate the conditions for static
equilibrium, some motion will result. Unlike a fluid, however, for a
granular
medium we  expect such motion to be in the form of a finite rearrangement rather
than a steady flow. Such a rearrangement will change the micro-texture of the
material, and thereby {\it alter the constitutive relation among stresses}. One
expects it to do so in such a way that the new (spatially inhomogeneous) constitutive relation 
becomes compatible with the imposed forces. 

Within this modeling approach, a granular
assembly is therefore able to support some, but not all, of the surface loads that would be supportable by an elastic medium.
Such models may therefore provide an
interesting paradigm for the behaviour of ``fragile matter" \cite{prl}.

\section{Experimental and numerical determination of the stress response function}
\label{response}
The stress `response function' to a localized overload is of prime interest
both from a fundamental point of view but also for many engineering applications.
On large scales, the extra stresses created by a house within the soil beneath it
are indeed related to this response function. Therefore, this problem has received
considerable theoretical attention in the engineering community, where, as mentioned
in the introduction, the granular material is often assumed to be a (possibly anisotropic) elastic
material. Note that an elliptic response function corresponds to a favorable case
for stability since the stresses are efficiently dispersed in space, whereas a
hyperbolic response function would lead to a rather localized stress field.

In spite of its importance, the response function of granular assemblies has
only very recently been measured experimentally \cite{manip2D,Reydellet,Chicago}. 
Various methods were used: one is a
direct quantitative measurement of the response at the bottom of a 3D packing
using a local stress probe based on the deformation of a hard membrane \cite{Reydellet}, 
another method is based on carbon paper imprints created by a
monodisperse 3D packing \cite{Chicago}. For 2D packing the photo-elastic
response of polymeric grains was used \cite{manip2D} in order to evidence
the inter-particle force path. This is a semi-quantitative method but it
allows to visualize directly the response in the bulk as well as the
topology of the path followed by the stress chains. Such an observation will be
used to built the theoretical proposition exposed in the following section.
These experimental efforts have lead so far to the following picture\footnote{%
Note however that a purely `diffusive' response function, scaling as
$\sqrt{H}$ as predicted by the `scalar' model, was reported in \cite{JER} for a very 
special `brick' packing.}:
for strongly disordered packings (for example by considering mixtures of grains
of very different sizes, or irregular grains such as natural sand), the
response profile on large length scales shows a single broad peak
\cite{manip2D,Reydellet}. This single hump response function was also
observed in numerical simulation
of 2D polygonal grains packing \cite{Moreau}. 

For well ordered packings however, the two peaks structure is rather
convincingly observed in two dimensions (a `ring' or three peaks in $3D$) \cite
{manip2D,Chicago,Bonamy}. These hyperbolic features are also in agreement
with a recent numerical simulation on (rather small) isostatic assemblies of
frictionless grains \cite{Witten,Head}, 
with the work reported in \cite{Eloy,Zucker} where beads are arranged on a 
regular triangular lattice but
where disorder is introduced by a finite friction coefficient. Similarly, two-peak
response can be observed for strongly anisotropic elastic networks \cite{Goldhirsch,Otto}.

Obtaining a precise experimental determination of the linear response function that can
be {\it quantitatively} compared to theoretical models is rather difficult:
the perturbation must be small enough not to disrupt the packing, but also
large enough to lead to a measurable signal. A very sensitive technique, based on a
lock-in detection of an oscillating perturbation, has allowed one to obtain precise
and reproducible results \cite{Reydellet,Reydellet2}. One finds unambiguously that the 
response function $G(r,H)$ to a perturbation
located at $r_0=z_0=0$ is in the case of sand layers {\it single-peaked}, with a width growing 
as the height $H$ of the layer. More precisely, the response function for different
heights can be rescaled onto a unique curve by plotting $H^2 G(r,H)$ as a function
of $r/H$. The factor $H^2$ is expected from
force conservation: the integral of the response function over the bottom plate
must be equal to the overload force $F$ for the total force to balance. One 
also finds that, like for sand-piles, the pressure
response profile depends on the way the granular layer was prepared -- its
`history': the value of the maximum of this response is roughly twice smaller
(and its width twice larger)
for a dense packing than for a loose one (see Fig.~\ref{fits}). 
The quantitative shape of the experimental response
function cannot be accounted for by a simple isotropic elasticity theory (see also the 
discussion in \cite{Tanguy}), but 
in the present geometry, anisotropic elasticity leaves three extra adimensional 
constants that can give a large freedom to fit the data (see \cite{Otto}).

\bfig[t!]
\bc
\epsfysize=6.2cm
\epsfbox{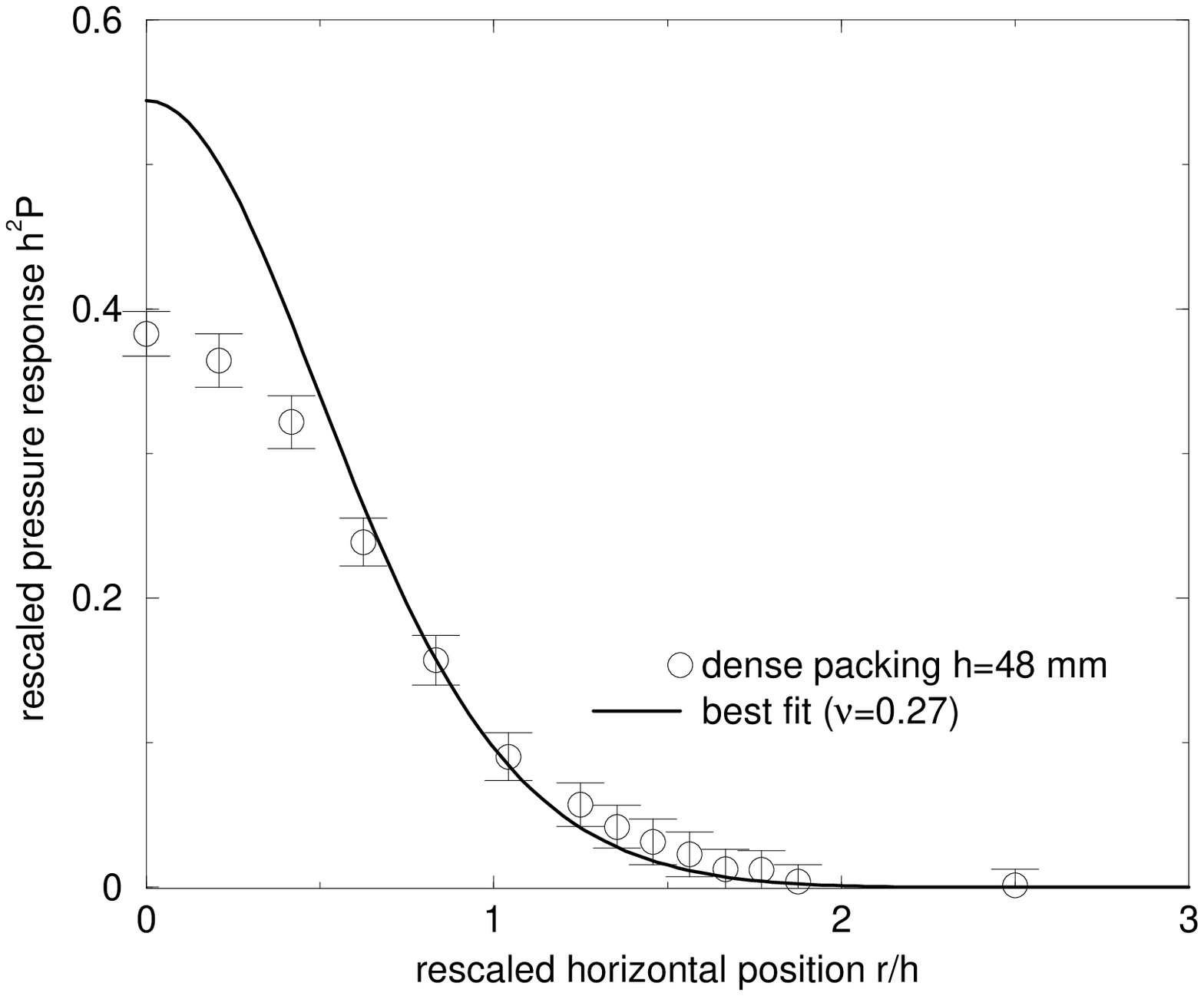}
\hfill
\epsfysize=6.2cm
\epsfbox{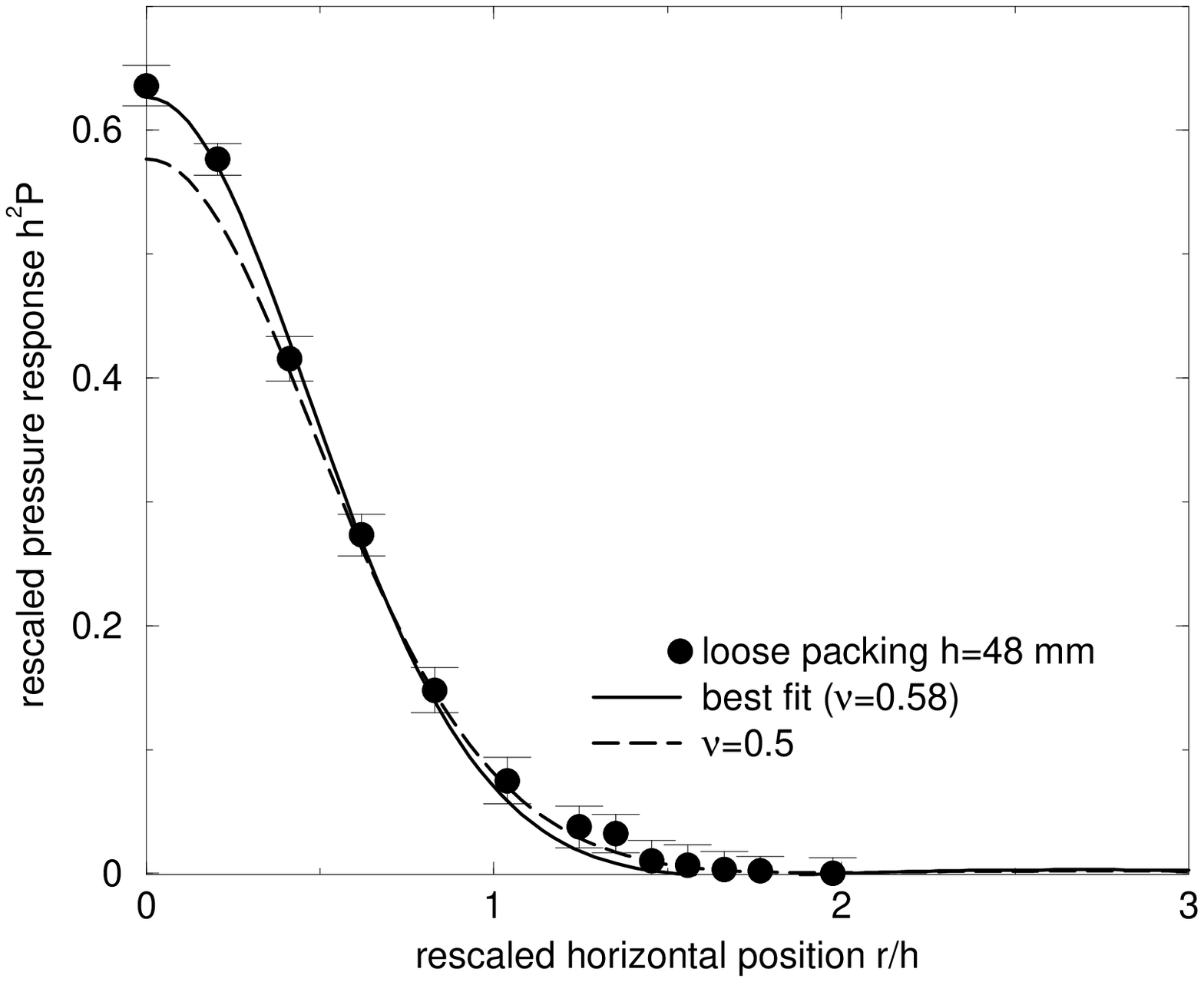}
\caption{\small Left: Fit of the `dense' packing data with the standard
elastic Green function. The agreement is not good at
all. Right: Fit of the loose packing data. The best fit value for $\nu$
exceeds the elastic bound $\nu \le \frac{1}{2}$. From \protect\cite{Reydellet2}.
\label{fits}}
\ec
\efig         

Therefore, the most precise experiments on strongly disordered packings seems to favor 
a single peak, elastic like response, rather than a double peak response that only 
appears in special conditions (ordered packings, or frictionless grains). We therefore
need to understand in details the r\^ole of randomness in hyperbolic wave equations, or 
in more physical terms, the large scale consequence of the fact that force chains are
`scattered' by packing irregularities. Is this sufficient to convert a two peak response 
function into a single peak, elastic like function? This is the subject of the
following sections.

\section{Force chains scattering I: weak disorder limit}

\subsection{A stochastic wave equation}

Provided that local conservation laws (those arising from
mechanical equilibrium)  are obeyed, many local rules for force transmission are
{\it a priori} compatible with the existence of contacts among rigid particles
\cite{Eloy,Socolar}.  Therefore, even if there is a definite mean relationship
among stresses at the meso-scale, one can
expect
randomness in the local transmission coefficients. The simplest model for
this and other sources of randomness is
to introduce a randomly varying `speed of light'
$c_0$. This could describe the fact that, for example, the parameter $p$
in the model of figure \ref{threeleg} can vary from grain to grain. In this situation
the two rays are still symmetric around the vertical direction, but with a random
opening angle. The situation where the bisecting line itself is random (i.e. when 
both the above opening angles $\psi_{\pm}$ are fluctuating) will be alluded to below.

This suggests the following stochastic wave equation for
stress propagation in two dimensions:
\be
\partial_{tt} \szz = \partial_{xx} \left[c_0^2(1+v(x,t))\ \szz\right]\label{RW}
\ee
where the random noise $v$ is assumed to be correlated as 
$\la v(x,t)v(x',t') \ra=\sigma^2 g_x(x-x') g_t(t-t')$. The correlation lengths 
$\ell_x$ and $\ell_t$
are again kept finite, and of the same order of magnitude.
In Fourier transform, this relation can also be written
$\la v(k,t)v(k',t') \ra=2\pi \sigma^2 \de(k+k') \tilde g_x(k) g_t(t-t')$. It
turns out that the final shape of the averaged response function depends on the
sign of $\tilde g_x(\La)$. In section \ref{ScalarII} we implicitly made the
choice $\tilde g_x(k)=1$, which corresponds to:
$g_x(x=0)=1/a$ and $g_x(x > 0)=0$. We will keep this choice for the following
calculations, but note that another form for $g_x$ could lead to
$\tilde g_x(\La)) <0$.  

In the following, $\szz$ will be again denoted by $w$. After a Fourier transform along $x$-axis,
we get, from equation (\ref{RW})
\be
\label{szz}
(\partial_{tt} + c_0^2k^2) w = \partial_t \rho -
c_0^2k^2 \int {dq \over 2\pi} w (q,t) v(k-q,t)
\ee
Note that the `source' term of this equation is now $\partial_t \rho$ rather
than $\rho$ itself. Therefore,
if we call $G$ the Green function (or propagator) of this equation
$G=\la \de w / \de \partial_t \rho \ra$; the response function
$R=\la \de w / \de \rho \ra$ of our system is now actually the
time derivative of $G$: $R(k,t)=\partial_t G(k,t)$.

The noiseless propagator $G_0$ is the solution of the ordinary wave equation
$(\partial_{tt}+c_0^2k^2)G_0(k,t-t')=\de (t-t')$ and can be easily calculated:
\be
\label{P0}
G_0(k,t) = {1 \over 2 \imath c_0 k} \left [e^{\imath c_0 k t}
- e^{-\imath c_0 k t} \right ] \theta (t)
\ee
which leads to the response function $R_0$
\be
\label{G0vectreal}
R_0(x,t) = {1 \over 2}
\left [ \de \left ( x-c_0 t \right ) + 
\de \left ( x+c_0 t \right ) \right ] \theta (t)
\ee
This last equation sums up one of the major results of the hyperbolic 
approach of  \cite{BCC,WCCB,FPA}:
in two dimensions, stress propagates
along two characteristic rays. (Note that the corresponding response function in three dimensions
reads $R_0(x,t) \propto (c_0^2 t^2 -x^2)^{-1/2}$ for $|x| < c_0 t$ and zero otherwise \cite{BCC}).
A relevant question is now to ask how these rays survive in the presence of disorder.
We will show that for weak disorder, the $\de$-peaks acquire a finite (diffusive) width,
and that the `speed of light' is renormalized to a lower value. Not surprisingly, the effect
of disorder can be described by an `optical index' $n>1$. For a strong disorder, however, 
we find (within 
a Gaussian approximation for the noise $v$) that the speed of light vanishes and becomes 
imaginary.
The `propagative' nature of the stress transmission disappears and the system might behave more 
like an
elastic body, in a sense clarified below.

\subsection{Calculation of the averaged response function}

One can again use Novikov's theorem in the present case if the noise is Gaussian and
short range correlated in time. However, the same results are again obtained within
the diagrammatic approach explained in section \ref{ScalarII}, which can be easily
transposed to the present case, and is more general. The propagator $G$ is a now
represented as a line, the source $\partial_t \rho$ a cross and the vertex meaning
$-c_0^2k^2\int{dq \over 2\pi}$. Within the {\sc mca}, the self-consistent equation
(analogous to equation (\ref{diagequa4}) in the scalar case) is:
\be
\label{H}
(\partial_{tt} +c_0^2k^2)H(k,t)=\delta(t) + \int_0^t dt' \Sigma_{\sc mca}(k,t') H(k,t-t')
\ee
where $H$ is defined by $G(k,t)=H(k,t)\theta(t)$, and the self-energy
$\Sigma_{\sc{mca}}$ given as
\be
\label{sevect}
\Sigma_{\sc{mca}} (k,t-t') =
2 \pi c_0^4 \sigma^2 k^2 \int {dq \over 2\pi} q^2 g_t (t-t')\tilde g_x(k-q) H(q,t-t')
\ee
Equation (\ref{H}) can be solved using a standard Laplace transform along
the $t$-axis ($E$ is the Laplace variable). Using the fact that $H(k,\tau)= \tau$ in
the limit where $\tau \to 0$, we find, for small $k,E$ (corresponding to scales $L$ such
that $\ell_x,\: \ell_t \ll L$): $H^{-1}(k,E)=E^2+\beta E+c_R^2k^2$, where
\bea
\label{coeff1}
c_R^2(k) & = & c_0^2 - {c_0^4 \sigma^2 \La^3\ell_t \over 6}\left 
(1-{3|k| \over 2\La} \right )
+ {\cal O}(k^2)\\
\label{coeff2}
\beta(k) & = & {c_0^4 \sigma^2 k^2 \La^3 \ell_t^2 \over 9} + {\cal O}(k^3)
\eea

We notice here that in the limit $\ell_t \to 0$, the effect of the randomness completely
disappears, as in the scalar model with the Ito convention. (Technically, this is due to the
fact that $G(k,t=0) \equiv 0$ in the present problem). In order to calculate the inverse
Laplace transform, we need to know the roots of the equation $H^{-1}(k,E) = 0$. This
leads to several phases, depending on the strength of the disorder. 

\vskip 0.5cm
$\bullet$ The weak disorder limit.
\vskip 0.5cm
For a weak disorder, $c_R^2(k)$ is always positive. We can then define $c_R=c_R(k=0)$.
As we will show now, $c_R$ is the shifted `cone' angle
along which stress propagates asymptotically. $c_R$ is a decreasing function of $\sigma$,
meaning that the peaks of
the response function get closer together as the disorder increases \footnote{As a technical 
remark, let us note that 
if $g_t=g_x$, the problem is symmetric in the change $x \to t$, $c^2(x,t) \to 1/c^2(x,t)$. 
It thus looks as if the cone should both narrow or widen, depending on the arbitrary choice
of $x$ and $t$. There is however no contradiction with the above calculation since we 
assumed that 
$v$ has zero mean, while $1/(1+v)-1$ has a positive mean value, of order $\sigma^2$}. 
For a critical 
value $\sigma=\sigma_c$, $c_R$ vanishes, and becomes
imaginary for stronger disorder.

In the limit of large $t$, the propagator reads: \be
\label{Pa}
G(k,t)= {1 \over c_R k} \sin \left [
c_R kt(1+\alpha |k|) \right ] e^{-\ga k^2t} \theta(t)
\ee
where the following constants have been introduced \footnote{Note that 
the sign of $\alpha$ is dictated by the sign of $\tilde g_x(\Lambda)$}:
\bea
\label{ctes1}
\alpha &=& {3 \over 4\La} \left ( {c_0^2 \over c_R^2} -1 \right ) \\
\label{ctes2}
\ga &=& {\beta(k) \over 2k^2} = {\sigma^2 \La^3 \ell_t^2 \over 18}
\eea
From equation (\ref{Pa}), the response function $R$, in the limit of small $k$ and large $t$,
is given by:
\be
\label{R}
R(k,t)= \cos \left [
c_R kt(1+\alpha |k|) \right ] e^{-\ga k^2 t} \theta(t)
\ee
or in the real space,
\be
\label{Gvw}
R(x,t)  = {1 \over 2\sqrt{4\pi|\gah|(t)}}
\Re \left\{ {e^{-\xi_+^2/ b} \over \sqrt{b}} 
\left [1-\Phi (-\imath {\xi_+ \over \sqrt{b}}) \right ] +
\sqrt{b}\: e^{-b \xi_-^2}
\left [1-\Phi (-\imath \sqrt{b} \xi_-) \right ] \right \}
\ee
\bfig[hbt]
\bc
\epsfysize=8cm
\epsfbox{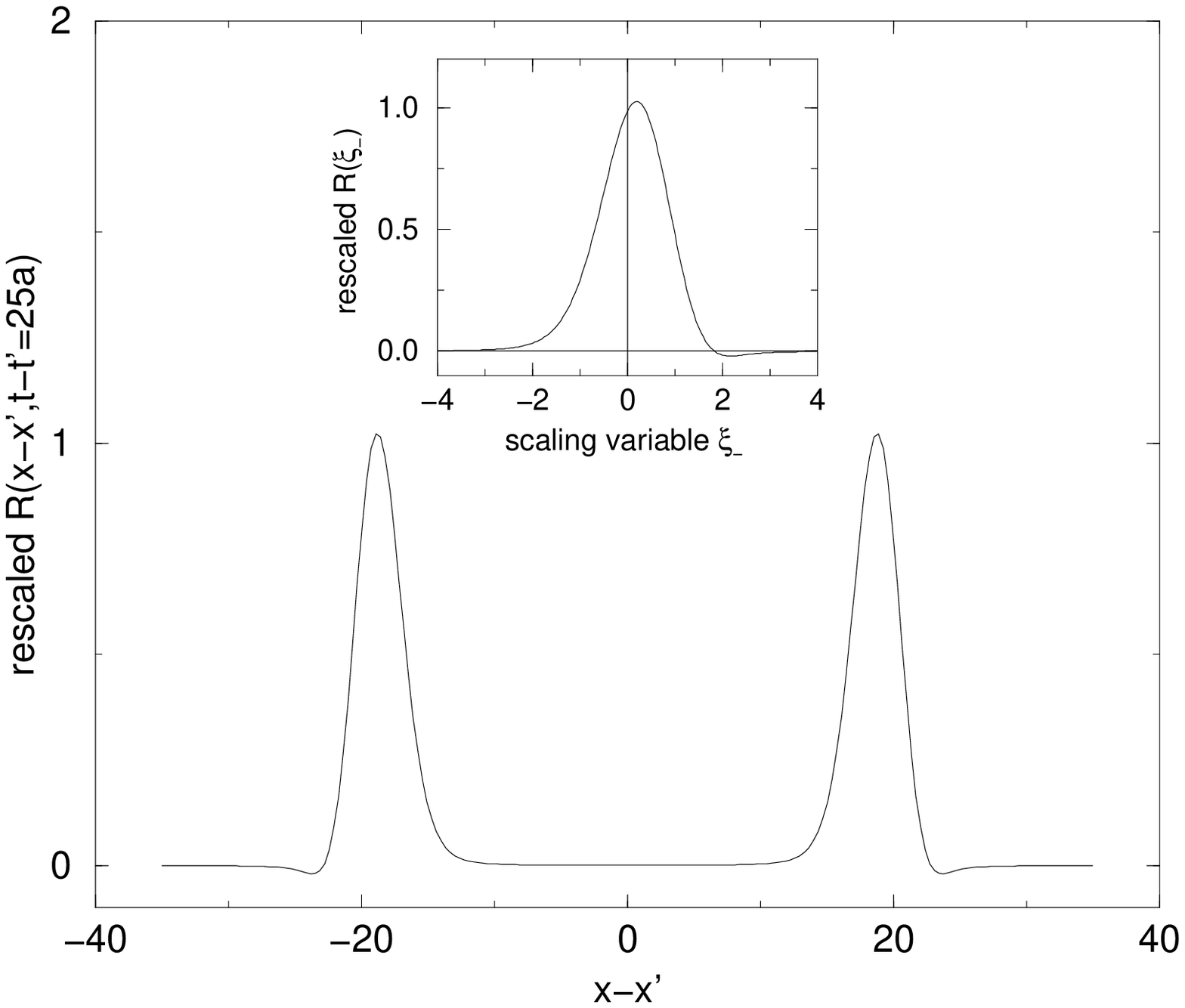}
\caption{\small Response function for weak disorder ($\sigma/\sigma_c \sim 0.32$).
The two curves have been rescaled by
the factor $2\left [4\pi|\gah|t \right]^{1/2}$.
The main graph shows the general double-peaked shape of the
response of the system when subjected to a peaked overload at $x=0,\ t=0$.
The inset gives details the right-hand peak as a function of the scaling
variable $\xi_-$. Note the asymmetry (for $\tilde g_x(\Lambda)>0$), compatible with
the results found in \protect\cite{Eloy}. Note also that the curve becomes
negative around $\xi_-=2$.
\label{Gwdfig}}
\ec
\efig
where the scaling variables $\xi_\pm$, measuring distances relative to the two peaks, 
are defined by
\be
\label{xipm}
\xi_\pm = {x \pm c_R t \over \sqrt{4|\gah|t}}
\ee
and where $\gah = \ga - \imath c_R \alpha$, 
$b = e^{\imath \arg{\hat{\ga}}}$. $\Phi$ is the standard error function.
Figure \ref{Gwdfig} shows $R$ as given by expression (\ref{Gvw}).
Interestingly, this propagator not only has a finite diffusive width $\propto \sqrt{t}$, 
but is also asymmetric around its maxima.
Surprisingly, and in sharp contrast to the scalar case discussed above, the response 
function becomes {\it negative} in certain intervals (although its integral is of course
equal to one because of weight conservation). This means that pushing on a given point 
actually reduces
the downward pressure on certain points. This can be interpreted as some kind of arching: 
increasing the
shear stress does affect the propagation of the vertical stress, and may indeed 
lead to a
reduction in its local value which is redistributed elsewhere. The un-averaged response function 
therefore necessarily takes negative
(and in fact rather large \cite{pre}) values. This is actually of crucial importance since this is
the source of fundamental `fragility' of granular matter to external
perturbations. Suppose indeed that as a result of the perturbation, a
grain receives a negative force larger than the preexisting vertical pressure.
This grain will then move and a local rearrangement of contacts will occur,
inducing a variation of $c_0(x,t)$ as to reduce the cause of the instability.
Thus, the stochastic wave
equation implicitly demands rules similar to those introduced, e.g. in \cite{CB}. The present
model, which is purely static, does not say what to do when a local rearrangement
occurs, but certainly suggests that small perturbations should induce such
rearrangements.  

It is interesting to note that this response function was numerically measured in \cite{Eloy}; its
shape is compatible with the above expression; in particular, the two peaks were 
found to be
asymmetric with a longer `tail' extending inwards, as we obtain here.
Note however that for $\tilde g_x(\La)<0$, the shape of the
peaks is reversed: the small dips are located inside the peaks and the longer
tail extends outwards. In a more recent work \cite{Zucker}, both the diffusive
spreading and the renormalisation of the wave velocity have been observed.

\vskip 0.5cm
$\bullet$ Critical disorder: The wave/diffusion transition.
\vskip 0.5cm
When the disorder is so strong that $c_R$ just vanishes, the roots of
$H^{-1}(k,E) = 0$ change nature, and so does the response function $R$.
The two peaks of the previous expression for $R$ merge together, while the width
becomes anomalously large ($\propto t^{2/3}$). In the asymptotic, large $t$, regime we obtain:
\be
\label{Pvs}
R(k,t) = \theta(t)\cos \left [ \lambda |k|^{3/2} t \right ] e^{-\ga k^2 t}
\ee
where the new constant $\lambda$ is defined by $\lambda=c_0\sqrt{3/2\La}$ and 
$\ga=c_0^2\ell_t/3$.

\vskip 0.5cm
$\bullet$ Strong disorder: The pseudo-elastic regime.
\vskip 0.5cm

For larger disorder still, one finds, within the {\sc mca} (which is exact for a Gaussian,
uncorrelated noise), that the renormalized value of $c_0^2$, $c_R^2$, becomes negative.
Upon a rescaling of $x$ as $\hat x = x/(ic_R)$, the effective equation on $\la \delta \szz \ra$
then becomes, on large length scales, Poisson's equation:
\be
\nabla^2  \la \delta \szz \ra = \partial_t \la \delta \rho \ra
\ee
which means that the stress propagation becomes somewhat similar to that in an elastic body, 
where stresses obey 
an elliptic equation of similar type \cite{LL}. In particular, the cone structure of stress
propagation, which is associated with the underlying, hyperbolic, wave equation finally 
disappears; the average response to a localized perturbation becomes a
broad `bump' of width comparable to the height of the pile. However, the above transition 
is possibly an artifact, due to the fact that $v$ is taken to be
Gaussian, which strictly speaking is not allowed, since the local value of $c_0^2$ should 
always
be positive. One can show for some other problems of the same type that a similar transition 
is artificially induced by
the Gaussian approximation when it cannot really exist on physical grounds. We need to 
address the strong disorder limit with different tools in order to discuss the large 
scale nature of the response function in this case: this will be discussed below. 

\subsection{Generalized wave equations}

It is tempting to generalize equations (\ref{equiF1}, \ref{equiF2}) and write the
most general linear equations governing the propagation of the forces
which are compatible with the (local) conservation rules. These equations
were first written by de Gennes \cite{PGDG}:
\bea
\label{equiF1bis}
\partial_t F_t + \partial_x \left[\eta'(x,t) F_x + \nu'(x,t) F_t \right] & = & \rho \\
\label{equiF2bis}
\partial_t F_x + \partial_x \left[\eta(x,t) F_t + \nu(x,t) F_x \right] & = & 0
\eea
Note that the terms $\nu,\,\nu'$ break the symmetry $x \to -x$, and exist whenever the
transfer rules in the three leg model are not symmetrical. Equations
(\ref{equiF1bis}, \ref{equiF2bis}) describe a situation where not only the opening 
angle of the cone can vary in space, but also its average orientation.

The same analytical techniques as above can  still be used. We shall only
discuss some special cases \footnote{To lowest order in perturbation theory,
the case where disorder is present in the four terms $\eta,\eta',\nu,\nu'$
simultaneously is very simply obtained by adding the corrections induced by each term 
taken individually.}:

$\circ$ Let us first set $\nu=\nu'=0$ and consider the case where $\eta'$ is random,
and $\eta$ fixed (and equal to $c_0^2$). Taking $\eta'(x,t)=\eta'_0 \ (1+v(x,t))$ with
the noise $v$ as above, one finds that the renormalized value of $\eta'$ is:
\be
\eta'_R = \eta_0' \left(1- {c_0^2 \eta_0' \sigma^2 \La^3\ell_t \over 6}\right)
\ee
Now, on large length scales, one must recover the continuum equilibrium equations
for the stress tensor, equations (\ref{equi1}, \ref{equi2}). The condition of zero 
torque requires
that the stress tensor is symmetric, and thus one must set 
$\eta'_R \equiv 1$, which imposes a relation between $\eta'_0$ and the amplitude
of the noise $\sigma$. Note that beyond a certain value of $\sigma$, this
relation can no longer be satisfied with a real $\eta_0'$. This again means that the
packing is unstable mechanically and will rearrange so as to reduce the disorder.

$\circ$ Another interesting 
class of models, which one can call `$\mu$ models', is such that:
$\eta=c_0^2, \eta'=1$, but $\mu(x,t)=c_0 v(x,t)$ and $\mu'=0$ or vice-versa.
These two cases yield identical results, namely, in the large $t$ limit:
\bea
\label{cas1et4}
R(k,t) & = & \cos \left ( c_0 k t \right ) e^{-\ga k^2 t} \theta(t) \\
\label{cas1et4'}
R_s(k,t) & = & -\imath c_0 \sin \left ( c_0 k t \right ) e^{-\ga k^2 t} \theta(t)
\eea
where $\ga = {c_0^2 \La \sigma^2 \over 4}$. Note that in these cases, the response peaks
acquire a finite diffusive width $\propto \sqrt{t}$, but the angle of the information cone
is unaffected by the disorder (i.e. $c_0$ is not renormalized).

$\circ$ Finally, there are special `symmetry' conditions
where the equations can be decoupled and reduced to two
`scalar' models. We will refer to this case as the `double scalar' model. This
occurs when $\mu=\mu'=c_0v_1(x,t)$ and $\eta'=\eta/c_0^2=1+v_2(x,t)$ where $v_1, v_2$ are
two different sets of noise. Let us define
$\sigma_\pm = c_0  F_t \pm F_x $, $x_\pm = x \mp c_0 t$ and $v_\pm = v_1 \pm v_2$, we then obtain:
\bea
\label{msc1}
\partial_t \sigma_+ & = & c_0 \rho -c_0 \ \partial_{x_+} [v_+ \sigma_+]\\
\label{msc2}
\partial_t \sigma_- & = & c_0 \rho -c_0 \ \partial_{x_-} [v_- \sigma_-]
\eea
showing that $\sigma_+$ and $\sigma_-$ decouple, each propagating along two rays, of `velocity' $\pm c_0$, 
plus a small noise $v_\pm$ which, as in the scalar case, generates a nonzero diffusion constant. 
The response functions for $\szz$ and $\sxz$ are thus again made of two diffusive peaks of width
$\propto \sqrt{t}$, centered in $x = \pm c_0 t$. Note  that by construction, this special form of disorder 
does not lead to negative vertical stresses.

A physically relevant question is to know how local stresses are distributed. We have seen above
that within a scalar approach, an exponential-like distribution (possibly of the type $\exp -w^\beta$, with
$\beta \geq 1$) is expected \cite{Liu,Copper}. One can wonder whether this exponential
distribution survives within a tensorial description, and what happens for very small stresses $w \to 0$.
Unfortunately, the full distribution can only be computed analytically for the `double scalar' 
model; but numerical results have also been obtained for the random {\sc BCC} model \cite{pre}
and confirm that the exponential tail holds in the strong disorder limit. 

In the `double scalar' limit, the histogram of the stress distribution is obtained trivially
by noting that since $\sigma_+=w_1$ and $\sigma_-=w_2$ travel along different paths, they are independent
random variables. Taking $c_0$ to be unity for simplicity, one thus finds:
\bea
\label{his1}
P(\szz) & = & \int dw_1 \int dw_2 P^*(w_1) P^*(w_2) \delta(\szz-\frac{w_1+w_2}{2}) \\
\label{his2}
P(\sxz) & = & \int dw_1 \int dw_2 P^*(w_1) P^*(w_2) \delta(\sxz-\frac{w_1-w_2}{2})
\eea
where $P^*$ is the distribution of weight pertaining to the scalar case, which, as 
mentioned above, depends on the specific form of the local disorder and on the 
discretisation procedure. In the strong disorder case which leads to equation (\ref{exptail}) 
[in the case $N=2$], we thus find that $P(\szz)$ is still decaying
exponentially (it is actually a $\Gamma$ distribution of parameter $2N$), although its
variance is
reduced by a factor $2$. For $N=2$, one simply gets
\bea
\label{his3}
P(\szz) & = & {8 \over 3} \szz^3 e^{-2\szz} \\
\label{his4}
P(\sxz) & = & \left ( |\sxz| + {1 \over 2} \right ) e^{-2|\sxz|}
\eea
The interest of the `double scalar' limit of the hyperbolic model is to show that the
exponential tail Eq. (\ref{exptail}) has indeed a certain degree of universality, and
is not restricted to the scalar model.

\section{Force chains scattering II: strong disorder limit}

\subsection{Introduction and numerical results}

\bfig[t!]
\bc
\epsfysize=7cm
\epsfbox{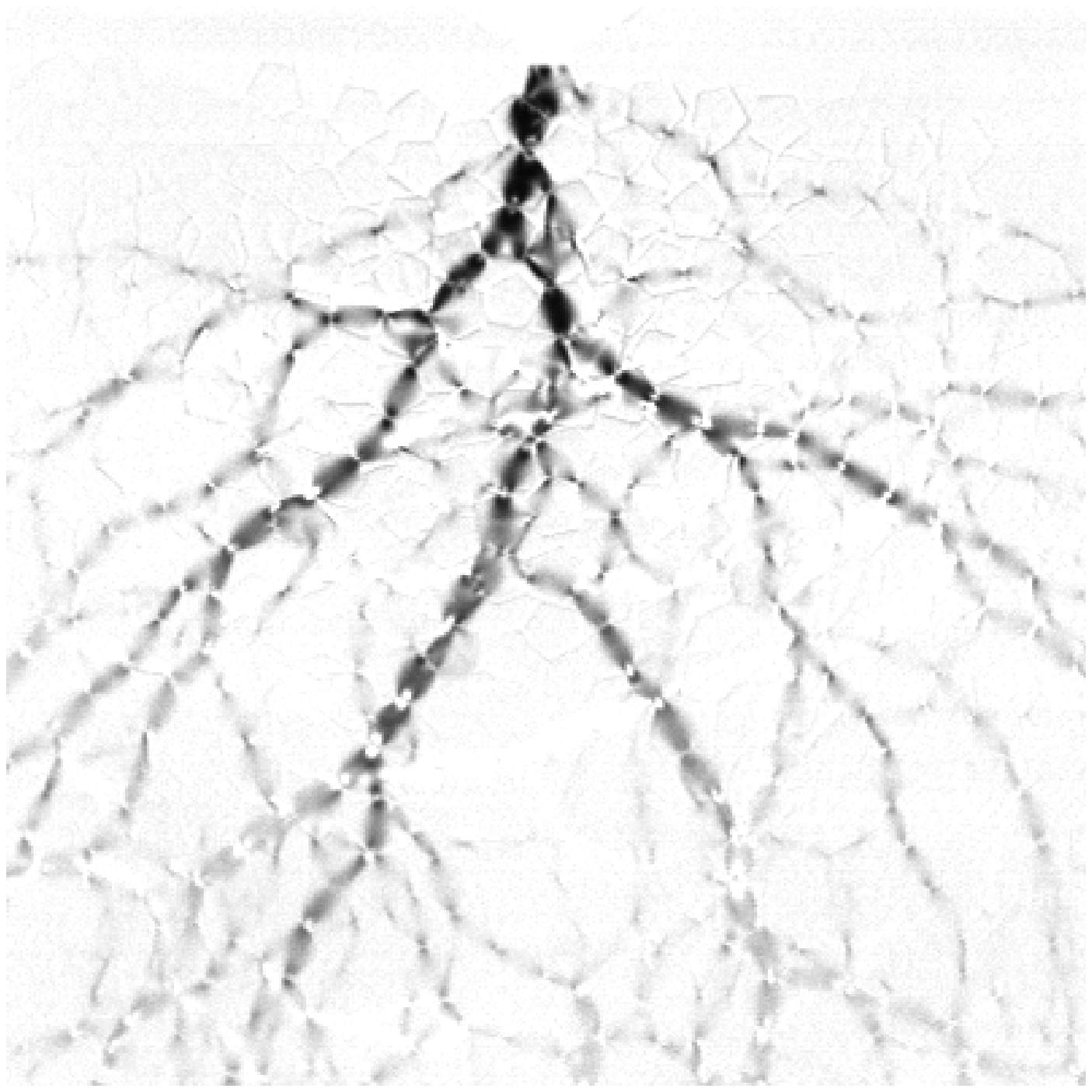}
\caption{Picture of the response force chains in a two-dimensional system of
grains subject to a vertical force imposed at the middle of the top surface.
To get this picture birefringent grains between inverse circular polarizers
have been used, and the intensity difference after and before the overloading
has been computed. This picture was obtained by R.P. Behringer and J. Geng.
\label{forcechainsexp}}
\ec
\efig

From pictures of photo-elastic grains, the network of inter-particle forces
propagating as a responses to a localized pressure was extracted \cite{Cambridge,manip2D} 
(see Figure \ref{forcechainsexp}). An
interpretation of such a picture can be given in terms of linear force
chains which tend to split upon meeting vacancies or packing defects \cite{bclo}.

As reviewed in the previous section, one can investigate perturbatively the role of disorder on
hyperbolic equations. In this case, the two peak structure of
the response function is preserved on large length scales, although the peaks
are diffusively broadened. This result is in qualitative agreement with the
numerical simulations \cite{Eloy,Head,Zucker}. An uncontrolled extrapolation to
strong disorder however suggests that the large scale equations might become
elliptic.

In order to investigate more quantitatively the strongly disordered regime
where force chains split and merge, one can study \cite{bclo} the following
model. If one of the force chains meets a defect (randomly distributed in space),
we split it into two new ones at random angle, which then propagate until another
defect (or the boundary) is reached. More precisely, a chain carrying a force $f$
in the direction $\vec n$ splits when meeting a defect into two forces $f_1$,
$f_2$ in the directions $\vec n_1$, $\vec n_2$ -- `\dy\ process'.  The two angles
$\theta_1$ and $\theta_2$ (between $\vec n$ and $\vec n_1$ and $\vec n_2$
respectively) are uniformly chosen between $0$ and some maximum splitting angle
$\theta_M$.  The local mechanical equilibrium imposes that the intensities $f_1$
and $f_2$ are such  that $f \vec n = f_1 \vec n_1+ f_2 \vec n_2$. Sometimes, two
(or more) force chains meet at the same defect -- `\uy\ process'. In this case, we
make them merge together. It is important to note that the positions of the defects
are fixed before starting the computation of the forces. This idea of a frozen
disorder is consistent with the experimental observation that when the local
overload is added on the top of the system, the forces are transmitted along the
chains originally created during the building of the packing. In other words, as
long as the applied force is not too large and compatible with the pre-existing
network of force chains, the geometry of the packing, and in particular the
contacts between grains, remains the same (but see the discussion in section~\ref{discussion}).

With these rules, realistic force networks can be created numerically \cite{bclo}. 
After averaging over many statistically identical samples,
one can obtain stress profiles for different heights $H$. One finds that, as $H$ increases, 
the vertical pressure
response profile evolves continuously from two well defined peaks to a single broad
one. It means that the hyperbolic behaviour is progressively erased by multiple
scattering. The width of the single peak is found to scale like $H$;
the scaling function is furthermore surprisingly close to the pure elastic
response of a semi-infinite two-dimensional medium. However, the numerical 
simulation was restricted to rather shallow (small $H$) samples. 

\subsection{A Boltzmann description of force chain splitting}

In order to understand analytically the above numerical results, we
write a Boltzmann equation for the probability density $P(f,\vec n,\vec r)$ of
finding an oriented force chain of intensity $f$ in the  direction $\vec n$ around
the point $\vec r$ \cite{bclo}. A very important point here is that force chains can
be oriented in reference to the boundary conditions (see the discussion in
\cite{Proc,Witten,bclo}).

For simplicity, we first neglect the chain `merging' process which leads to 
a more complicated non linear Boltzmann equation (its influence is not fully
understood yet and will be discussed below). We also assume that the splitting is
symmetric, i.e that $\vec n\cdot\vec n_1=\vec n\cdot\vec n_2 \equiv \cos \theta$,
so that $f_1=f_2=f/2\cos\theta$. Assuming a uniform density of defects, the
probability distribution $P(f,\vec n,\vec r)$ obeys the following general
equation:
\be
P(f,\vec n,\vec r+\vec n \, dr) =
(1-\frac{dr}{\lambda}) P(f,\vec n,\vec r) \, +
2 \frac{dr}{\lambda} \! \int \!\! d\vec n' \!\! \int \!\! df'
P(f',\vec n',\vec r) \Psi(\vec n',\vec n)
\, \delta \! \left(f-\frac{f'}{2\cos\theta}\right)\!,
\label{Boltzmann}
\ee
where $\lambda$ is equal to the `mean free path' of force chains, and is of
order $1/(\rho_d a^{d-1})$ in dimension $d$. The above equation means
the following: since a chain of grains can only transport  a force parallel
to itself \cite{prl}, the direction of the force $\vec n$  also gives the local
direction of the chain. Between $\vec r$ and $\vec r+\vec n \, dr$, the chain
can either  carry on undisturbed, or be scattered. The second term on the
right hand side  therefore  gives the probability that a force chain initially
in direction $\vec n'$ is scattered in direction $\vec n$. This occurs with
a probability $\frac{dr}{\lambda} \Psi(\vec n',\vec n)$, where
$\Psi$ is the scattering cross section, which we will assume to depend
only on the scattering angle $\theta$, for example a uniform distribution between
$-\theta_M$ and $+\theta_M$. The $\delta$-function ensures force conservation
and the  factor two comes from the counting of the two possible outgoing
force chains. Let us now multiply equation (\ref{Boltzmann}) by $f$ and integrate
over $f$. This leads to an equation for the local average force per unit volume
in the direction $\vec n$, that we will denote $F(\vec n,\vec r)$. This equation
reads:
\be
\lambda \, \vec n \cdot \! \vec \nabla_r F(\vec n,\vec r) =
- F(\vec n,\vec r) \, +
\int \!\! d\vec n' \, \frac{F(\vec n',\vec r)}{\vec n \cdot \vec n'} \,
\Psi(\vec n',\vec n) + \frac{\lambda}{a} \, \vec n \cdot \vec F_0(\vec r),
\label{SM}
\ee
where we have added the contribution of an external body force density $\vec
F_0(\vec  r)$, and $a$ is the size of a defect or of a grain. This
equation is the so-called Schwarschild-Milne equation for radiative  transfer,
describing the evolution of light intensity in a turbid medium \cite{Theo}. We
now introduce some angular averages of $F(\vec n,\vec r)$ that have an
immediate physical  interpretation:
\bea
p(\vec r)                       & = &
a   \! \int \!\! d\Omega \, F(\vec n,\vec r) \\
J_\alpha(\vec r)                & = &
a   \! \int \!\! d\Omega \ n_\alpha \, F(\vec n,\vec r) \\
\sigma_{\alpha \beta}(\vec r)   & = &
a d \! \int \!\! d\Omega \ n_\alpha  n_\beta \, F(\vec n,\vec r),
\eea
where $\int \!\! d\Omega$ is the normalized integral over the unit sphere, that
is introduced for a correct interpretation of $\sigma$ (see equation (\ref{second})
below). As will be clear from the following, $\vec J$ is the local
average force chain intensity per unit surface, $\sigma$ is the stress tensor.
Since $\vec n^2=1$, one finds that Tr $\sigma=dp$, and therefore $p$ is the
isostatic pressure. Note that $\vec J$ is not the average local force, which
is always zero in equilibrium. The fact that $\vec J \neq \vec 0$ comes from the
possibility of \emph{orienting} the force chains.

We now integrate over the unit sphere equation (\ref{SM}) after multiplying it by
different powers of $n_\alpha$. Using the fact that $\Psi(\vec n',\vec n)$
only depends on $\vec n \cdot \vec n'$, a direct  integration leads to:
\be
\lambda \, \vec \nabla \! \cdot \vec J = (a_0-1) p,
\label{first}
\ee
where $a_0$ is called the `albedo' in the context of light scattering
\cite{Theo}, and reads:
\be
a_0 \equiv \int \!\! d\vec n \,
\frac{\Psi(\vec n',\vec n)}{\vec n \cdot \vec n'} \geq 1.
\ee
A second set of equations can be obtained by multiplying by $n_\alpha$ and
integrating. Using the fact that $\int \!\! d\vec n \, \Psi(\vec n',\vec n) = 1$,
it is easy to show that
\be
\int \!\! d\vec n \,\vec n \, \frac{\Psi(\vec n',\vec n)}{\vec n \cdot \vec
n'} =
 \vec n'.
\ee
Therefore, the resulting equation is nothing but the usual mechanical equilibrium
relation:
\be
\nabla_\beta \sigma_{\alpha \beta} = F_{0\alpha}.
\label{second}
\ee
This relation in fact only reflects the local balance of forces chains.
Now we multiply equation (\ref{SM}) by $n_\alpha n_\beta$ and again integrate. A
priori, this introduces a new three index tensor. In order to close the
equation, we now  make the an assumption that is usually made in the context
of light diffusion, that on large scales the force intensity becomes nearly
isotropic \cite{Theo}. In this case, it is justified to expand $F(\vec n,\vec
r)$ in angular harmonics and to keep only the first  terms:
\be
a \, F(\vec n,\vec r) = p(\vec r)+  d\vec n \cdot \vec J(\vec r) + ...
\ee
Within this expansion, we finally obtain a
`constitutive' relation between $\sigma_{\alpha \beta}$ and the vector
$\vec J$. We find:
\be
\left.\sigma_{\alpha \beta}\right|_{\alpha \neq \beta}=
\frac{\lambda d^2K_1}{\left(\frac{da_2-a_0}{d-1}-1\right)} \left(\nabla_\alpha
J_\beta +\nabla_\beta J_\alpha \right),
\ee
and
\be
\sigma_{\alpha \alpha} =
\frac{\lambda d}{\left(\frac{da_2-a_0}{d-1}-1\right)}
\left[ 2d K_1 \, \nabla_\alpha J_\alpha \, +
\left(dK_1-\frac{a_0-a_2}{(d-1)(a_0-1)} \right)
\vec \nabla \! \cdot \vec J \right]
\ee
with $K_1=1/d(2+d)$ and $a_2 = \int \!\! d\vec n \, (\vec n \cdot \vec n')
\Psi(\vec n',\vec n)$. 
Rather surprisingly, these equations have exactly the
canonical form of elasticity theory, provided one identifies the vector
$\vec J$ with the local displacement, up to a multiplicative factor.

The above stress equations are rather non-trivial because no displacement field is introduced
in the above derivation (nor in the numerical model): elastic-like equations
are found in a stress-only model. The basic assumption is the existence of
local force chains, which have a well defined identity over several grain
sizes $a$, such that $a \ll \lambda$: this is the condition under which the
above Boltzmann description of the force chain scattering is justified.

Since the above equations are formally identical to those of classical
elasticity, one can show that $\nabla^2 p = 0$, and $\nabla^4 \vec J = 0$ \cite{LL}.
One can therefore in principle compute the response function $G(\vec r)$ to a
localized force at $\vec r_0=0$ in the $z$ direction, which is given by the 
standard (one peak) elastic Green's function. But note however that although
the above equations are formally those of classical elasticity, there is one
crucial difference coming from the fact that $(\sab)$ and $\vec J$ are not
independent since they are both related to the same underlying quantity
$F(\vec n,\vec r)$. This is a very important difference, which appears, for example
in the choice of boundary conditions on $\cal B$ that determines 
$\left. P(f,\vec n,\vec r)\right|_{\cal B}$.

\subsection{The role of chain merging}

We have mentioned above the fact that the numerical simulation of the 
full chain scattering model (including both splitting and merging) was
restricted to small depths. Similarly, neglecting chain merging altogether
cannot be correct on large length scales, since the number of force
chains would diverge, leading to an infinite number of force chains with
infinitesimal intensity. As shown in \cite{ssc}, the above linearized theory
is in fact unstable, and chain merging play a crucial role to make the
theory mathematically consistent.

On the other hand, chain merging leads to a non-linear Boltzmann equation
which in the general case cannot be solved. Only special cases have, up to
now, been amenable to an analytic treatment. For example, if one insists that
the force chains can only take six directions a $60^o$ degrees, 
with $120^o$ degree chain splitting and chain merging, the amplitude of the force chains
is constant, simply because $2 \cos 60 = 1$. In this case, the full probability
distribution  $P(f,\vec n,\vec r)$ boils down to six functions $p_n(\vec r)$, describing the
probability for a force chain to be in one of the six available directions.
The result of the full analysis is that, quite surprisingly, the 
response function evolves back to a two-peak, hyperbolic structure on
large length scales! Whether this is due to the particular structure of 
the model is not yet settled; preliminary results on the eight-fold model
\cite{bcso} suggests that there might actually be a {\it transition} between a 
hyperbolic response for sufficiently anisotropic scattering and an
elliptic response for isotropic situations. (see \cite{Socolarnew} for a
nice recent discussion). The existence of such a 
transition would perhaps, as we discuss now, allow one to reconcile the
apparently contradictory experimental, numerical and theoretical results.

\section{Statics of granular materials: concluding remarks and open questions}
\label{discussion}
Let us try to summarize the theoretical situation as follows. If the grains
are frictionless, then packings are generically isostatic. If these packings
are furthermore `sufficiently' anisotropic, then the construction of the
stress everywhere in the packing can be performed by `propagating' the
stress from one boundary towards the interior of the packing, as one expects 
for hyperbolic equations. This situation is well captured by the three-leg
model introduced above, that indeed leads to hyperbolic equations on large
length scales. Correspondingly, numerical simulations of the response function 
in {\it anisotropic} packings of frictionless beads indeed display the 
expected diffusively broadened two-peak structure, at least for the modest sizes 
that were simulated \cite{Head}.

However, not all isostatic packings are characterized by a two-peak response 
function. For example, using {\it isotropic} isostatic packings built by
J.N. Roux, Ph. Claudin and A. Ayadim \cite{Claudin} have found that the average 
response function 
is a single hump, qualitatively similar to the elliptic-like
response observed experimentally and to the result of the chain splitting model. 
An example of such packings is shown in Fig.~\ref{resglav}, together with the local force 
chains. This picture makes it obvious that the force chains are strongly scattered, 
and lose their `coherence', which makes it indeed plausible that the two-peak structure 
is destroyed. Again, some degree of anisotropy seems to be required to preserve
hyperbolicity. Note that 
in the numerical determination of the response function, the applied force 
perturbation $\delta f$ is infinitesimal and does not induce any rearrangement of the initial
structure. 

\bfig[t!]
\bc
\epsfysize=7cm
\epsfbox{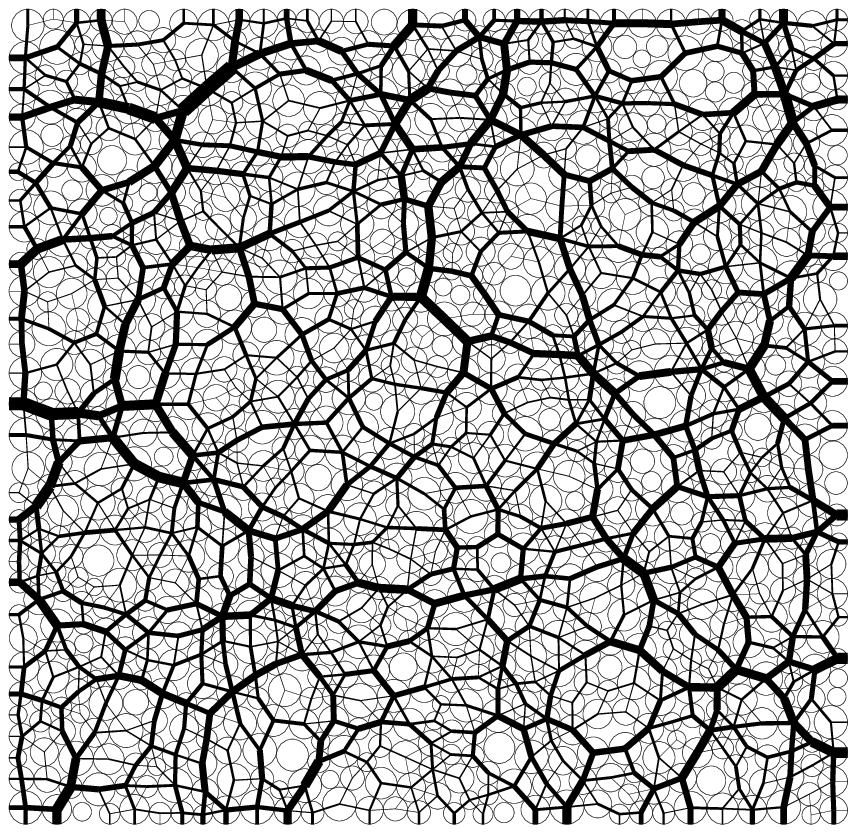}
\caption{Force chains in an isotropic, isostatic packing of frictionless disks.
The average response function in this case has a single peak. Courtesy of Ph. Claudin.}\label{resglav}
\ec
\efig

This remark is important since {\it if} rearrangements are allowed, the 
response has been argued in \cite{Roux} to be, on general grounds, elliptic.
Moreover, one expects that for any small perturbation, a sufficiently 
large assembly of frictionless grains will rearrange \cite{RouxPRL}. Therefore,
the order in which the limits $H \to \infty$ and $\delta f \to 0$ are taken 
is physically relevant \cite{RouxC,HeadC}, and one might expect that even  
anisotropic structures such as those studied in \cite{Head} should destabilize 
under a small perturbation for large enough $H$. It is not clear which of the 
limits is relevant in the experimental conditions of \cite{Reydellet}; even
if the applied force is very small and much care has been devoted to perturb as weakly as possible the packing, it might well be that these experiments 
are not in the infinitesimal limit. 

Conversely, is isostaticity needed to obtain hyperbolicity? The numerical 
simulations of \cite{Eloy,Zucker}, where an hyperbolic response function
for anisotropic ordered packing of grains with friction is observed, show
 that this is not the case. Similarly, the experimental determination of hyperbolic like response function in ordered lattices \cite{Chicago,manip2D} shows that, 
as suggested by the above theoretical analysis, weak disorder does not 
suppress the hyperbolic nature of the force propagation. It would be
extremely important, in this context, to confirm theoretically the existence
of a true `hyperbolic-elliptic' phase transition in simplified models of
force chain scattering. 

More formal approaches have also been advocated in \cite{Grinev,Ball} to 
try to establish some closure relations between the components of the
(coarse-grained) stress tensor starting from the equilibrium force balance
at the grain level. Tkachenko and Witten \cite{Witten} have insisted on the equivalence between any macroscopic linear relation between the components of the
stress tensor and the existence of `floppy' modes, i.e. zero energy 
deformation modes (see also \cite{Moukarzel}). Such floppy modes can exist in certain elastic spring 
networks, such as a square lattice which can deform at no cost along its
diagonal. One can indeed show that the large scale elasticity equations 
in this case are characterized by coefficients such that the marginal 
condition $r^2=t$ (see Eq.~(\ref{rootX}) above) is satisfied, so that the response 
function becomes the sum of two delta peaks \cite{Otto}. Following this line of 
thought, a natural conjecture is that if a system has some {\it extended}
floppy modes, then its large scale response will be hyperbolic. Conversely,
if floppy modes are all localized, then the response is locally hyperbolic, 
but the strong force chain scattering disrupts the long range propagation and
the response becomes elliptic like, as suggested by the force chain scattering model
discussed above. We hope that these issues will be clarified
in the near future; a particularly important point is to establish precisely, 
starting from a microscopic
description {\`a la} Edwards, which kind of large scale static equation emerge and under 
what conditions it is hyperbolic/elliptic, and the value of the parameters entering these equations.

\section{Glassy dynamics in granular media: a brief survey}
\label{dynamics}
\subsection{Slow compaction}

We have spent quite a long time on the static properties of granular media. The 
dynamics of {\it weakly tapped} assemblies of grains is also an extremely interesting
subject, in particular in relation with the dynamics of other glassy 
systems, such as super-cooled liquids, colloidal glasses and foams. The most obvious experiment 
(with important industrial applications) is that of compaction under tapping. The basic 
experiment consists in studying the volume occupied by a large number of grains when the 
container is `tapped', i.e. subject to a periodic acceleration of a certain amplitude. The
packing, initially very loose, progressively compacts. However, the compaction process slows 
down with time, and the decay of the volume is very far from being a simple exponential. 
Experiments have shown that the decay of the volume towards its asymptotic value 
can be satisfactorily fitted by an inverse logarithm of time, or by a stretched exponential,
both forms being typical of relaxation processes in glasses, where one can also study the time
dependence of the volume (or of the energy) as the sample relaxes after a temperature quench.
The parameters of these fits depend on the tapping amplitude -- stronger tapping obviously leads to a 
faster compaction. 

More complex experimental protocols have also been tested.
For example, one can change, in the course of the experiment, the amplitude of the 
tapping and reveal
interesting memory effects, again similar to those found in glasses and spin-glasses. 
The now classic 
experiment \cite{Nowak} is to start from a loose sample and increase slowly the tapping amplitude $\Gamma$, 
in such 
a way that for each tapping amplitude the density $\rho$ appears to reach a saturation value. 
One finds 
that $\rho(\Gamma)$ increases with $\Gamma$. When $\Gamma$ is reduced back to zero, the 
density keeps a
high value, revealing a kind of irreversibility that also appears in spin-glasses under 
a magnetic field:
when the temperature is increased the (zero field cooled) magnetization increases, but 
does not follow the
same path on the way back. The temperature at which the two branches separate is the 
spin-glass transition 
temperature (which, if defined in this way, weakly depends on the heating/cooling rate). 
Similarly, there
appears to be a tapping amplitude beyond which the two density branches meet. 
The high magnetization
(field cooled) branch, as the high density branch in the granular system, {\it is} reversible. 
As in
spin-glasses, the low density branch is in fact out-of-equilibrium, but the convergence 
of the density 
(magnetization) towards its equilibrium value is much too slow to be observed. 

In the first stage of another type of experiment \cite{Josserand}, aimed at revealing 
`memory effects', 
one taps the system with three different amplitudes -- say weak, moderate and 
strong -- during a time chosen such as to reach a certain density, identical in the
three cases. In the
second stage of the experiment, the tapping amplitude is then chosen to be moderate, 
and the density 
just after the amplitude `jump' is observed. If the state of the system was only described 
by the density, 
the evolution of the density after the jump should be identical for all three situations, 
and follow the `moderate' 
curve, which is taken as the reference. This is not the case:
the weakly tapped system first has to {\it dilate} before it is able to resume its compaction, 
whereas the strongly tapped
system compacts faster than the reference system just after the jump. This shows that the 
configurations reached under 
stronger tapping are in a sense easier to compact further. This experiment therefore indicates 
that some further `hidden' 
observables are needed to describe the large scale evolution of the system. 
We shall expand on this below, but want
to note that a very similar effect is known in glasses as the Kovacs (or memory) effect. 
In this case, one prepares 
the system at a given temperature $T_2 < T_1$ and waits until the volume has reached the 
equilibrium volume at $T_1$.
Then one raises the temperature to $T_1$. If the volume was the only relevant quantity, 
one should not see any 
further evolution since this volume is already at its equilibrium value. 
Again, this is not what Kovacs first observed in polymeric glasses \cite{Kovacs}:
the volume has to increase back to a maximum before decreasing again towards its 
equilibrium value. A similar
effect was also reported in numerical simulations of spin-glasses \cite{BB02} and of 
Lennard-Jones 
systems \cite{Bertininprep}, and several simple
models are now known to reproduce this effect (see below).

We also want to mention the very interesting experiment of d'Anna et al. \cite{dAnna}, 
where a torsion 
oscillator is immersed in a vibrated granular assembly. The observable is the angle 
$\theta$ made by the 
oscillator with an arbitrary axis. The results of \cite{dAnna} are that (a) the 
angle $\theta$ performs a
random walk in time: $\langle [\theta(t+\tau)-\theta(t)]^2 \rangle = D(\Gamma) \tau$ 
and (b) the angular 
diffusion constant $D(\Gamma)$ appears to vanish as the amplitude of the tapping $\Gamma$
goes to zero in a way 
that recalls the `super-activated' Vogel-Fulcher law in glasses.

\subsection{Self-inhibitory dynamics and dynamical heterogeneities}

\subsubsection{Non exponential relaxation}

We shall call $\rho^*$ the maximum compacity that can be reached in a tapping experiment. 
For $\Gamma \to 0$, this 
corresponds to the so-called `random close packing' configuration. 
[It is not entirely clear how meaningful this notion 
really is, but from an empirical point of view, any reasonable extrapolation of the accessible dynamics seems to converge 
to a density which is not the HCP density (in the case of identical spheres).] 
Correspondingly, we will call the difference 
between $\rho^*$ and the average density $\rho$ the `free volume' fraction, $\Phi$. 
The simplest relaxation equation for $\Phi$ is obviously
a rate equation:
\be
\frac{d\Phi}{dt} = - \gamma \Phi.
\ee
If the decay rate $\gamma$ is independent of the free-volume itself, the decay of $\Phi$ 
is obviously a 
single exponential. However, it is intuitively clear that the dynamics is self inhibitory, 
in the sense that it
is the free volume itself that allows further compaction. Therefore, one expects that 
$\gamma$ vanishes as $\Phi \to 0$.
Assuming a power-law dependence $\gamma \sim \Phi^\beta$, with $\beta > 0$, one obtains 
a power-law relaxation for 
long times:
\be
\Phi \sim (t+t_0)^{-1/\beta}.
\ee
Let us assume a simple model where particles have a volume $\upsilon$, and mobile 
holes have fixed `quantum' volume $\upsilon_0$.
For a particle to be able to move, we require that $n^*$ holes meet in a volume $\upsilon$, 
such that the volume
of one particle is liberated, i.e. $n^* \upsilon_0 =\upsilon$. If the dynamics is sufficiently 
chaotic, it is
reasonable to assume a Poisson distribution for the holes. Therefore, the probability for a 
particle to move is  
simply $\Phi^{n^*}$, leading to $\beta=n^*$. If the number of holes $n^*$ needed to move one 
particle is large, one can 
approximate the long time behaviour of $\Phi$ as a logarithm:
\be
\Phi \sim t^{-1/\beta} \approx \frac{1}{1+\frac{1}{n^*} \log t},
\ee
a form often advocated to fit the slow relaxation of the volume in glasses or granular 
media \cite{Struik,Nowak} (but see also \cite{philippe}, where the role of
convection is discussed, and \cite{Viot} for a simple soluble model).

\subsubsection{Dynamical heterogeneities}

If one assumes that holes cannot spontaneously appear, it is clear that the dynamics is 
necessarily spatially 
heterogeneous: regions rich in holes, where dynamics is locally fast, appear only to the 
detriment of regions poor
in holes, where the system is jammed. This argument can be somewhat formalized to suggest 
that the geometry of the `fast' 
objects is non trivial. Call $\Phi + \phi(\vec r,t)$ the coarse-grained density of
free volume (`holes') 
around point $\vec r$ at time $t$,
such that the space average of $\phi(\vec r,t)$ is equal to $0$. [Note that all `voids' do not necessarily contribute to the {\it free} volume].
Far from the boundaries of the sample where the holes 
can disappear, one can write a Langevin equation for $\phi(\vec r,t)$ which reads 
\cite{Dean}:
\be\label{fastdif}
\frac{\partial \phi}{\partial t} = D \nabla^2 \phi + \vec \nabla \cdot [\sqrt{\Phi+\phi} 
\,  \vec \xi(t)],
\ee 
where $D$ is the (fast) diffusion constant of the holes, $\vec \xi$ is a Gaussian white noise 
in space and time of variance 
equal to $D$, and we have neglected the interaction between the 
holes. In the above equation, we have supposed that the free-volume is locally conserved. 
Assuming that $\phi$ is small leads to the result that $\phi$ itself is a Gaussian 
field with a correlation
function given (in three dimensions) by:\footnote{If one considers that `holes' can be converted
into voids and vice-versa, such that $\phi$ is only conserved on average, the statistics 
of the $\phi$ field is altered and becomes that of an elastic membrane subject to thermal
fluctuations.}
\be
\langle \phi(\vec r,t) \phi(\vec r',t+\tau) \rangle= 
\frac{\Phi}{2} \frac{1}{(4 \pi D\tau)^{3/2}} 
\exp\left(-\frac{\left(\vec r-\vec r'\right)^2}{4D\tau}\right) \qquad (\tau \geq 0).
\ee
Note that in the limit $\tau \to 0$, the field $\phi$ is delta-correlated in space; 
some short scale cut-off 
of the order of the grain size $a$ would be needed to regularize this behaviour.
Now, if one insists that the material particles (or grains) can only move if the local 
density of holes is larger than
a certain threshold $\Phi_c$, the diffusion equation for, say, the density ${\rho}(\vec r,t)$ of slow tracer particles reads (see also \cite{Long} for similar ideas):
\be\label{slowdif}
\frac{\partial {\rho}}{\partial t} = D_0  \vec \nabla \cdot \left[\Theta(\Phi+\phi-\Phi_c) 
\vec \nabla 
{\overline \rho}\right].
\ee
Although a more careful solution of the coupled equations 
(\ref{fastdif},~\ref{slowdif}) 
is required to confirm the
following conclusions, a qualitative analysis of the problem leads to the following picture 
for glassy dynamics, 
which relates slow, self-inhibitory dynamics and dynamical heterogeneities: 
\begin{itemize}

\item The probability that an elementary region of space is active is, for small $\Phi \ll \Phi_c$, 
proportional 
to $\exp(-C\Phi_c^2/\Phi)$, where $C$ is a numerical coefficient. Therefore, the large 
scale diffusion coefficient 
$D_R$ of the particles vanishes as:
\be\label{VF}
D_R \sim D_0 \exp\left(-\frac{C\Phi_c^2}{\Phi}\right) = D_0 \exp\left(-\frac{C\Phi_c^2}
{\rho^*-\rho}\right),
\ee
i.e. {\`a la} Vogel-Fulcher. This argument is actually the analogue of the classic 
free-volume argument leading to the 
Vogel-Fulcher law in glasses, and extended to granular media by Boutreux and De Gennes 
\cite{Boutreux}.

\item The fast regions are delimited by the contour lines of a correlated 
Gaussian field. This has 
interesting consequences: for example, one expects, in the short-ranged conserved case, the active regions to be lattice 
animals of fractal dimension $d_f=2$
in three dimension (and $d_f \approx 1.56$ in two dimensions). The value $d_f=2$ 
turns out to be in agreement with the experimental determination of Weeks et al.
on dense, three dimensional colloidal glasses \cite{Weeks}. It would be interesting to 
measure other geometrical
characteristics of the fast clusters to confirm that these are indeed lattice
animals (see e.g. the discussion in \cite{Lamarcq}). Interestingly, the connection 
between glassy dynamics and the contour lines of a dynamic random field was also
suggested in \cite{Chamon}, using rather different arguments. 

\end{itemize}

It would be most interesting to make these statements more precise in the context of a 
specific model. Very promising
steps in that directions have been made recently in dynamically constrained models \cite{Garrahan,Biroli,Ritort-Sollich}.

As emphasized by Boutreux and De Gennes, a Vogel-Fulcher law such as Eq.~(\ref{VF}) also
leads to a logarithmic 
relaxation of the density. Indeed, writing:
\be
\frac{d\Phi}{dt} = -\gamma_0 \exp\left(-\frac{C\Phi_c^2}{\Phi}\right) \Phi,
\ee
leads at large times to:
\be
\Phi \approx \frac{C\Phi_c^2}{\log t}.
\ee

\subsubsection{Another point of view: Edwards postulate}

The above Vogel-Fulcher law can also be understood using Edwards' analogy with thermodynamics. 
Assume again that all blocked states are equiprobable. Then the states that are most likely to
be observed are the most numerous ones: this is the essence of statistical thermodynamics. 
Imagine for example that a container is divided by a movable piston, with grains in each 
compartments.
The volume of the two compartments are $V_1$ and $V_2$, with $V_1+V_2=V$. The Edwards 
entropy ${\cal S}$ is the logarithm of the 
number of blocked states for a given overall volume. The total entropy of the container is:
\be
{\cal S}_T = {\cal S}_1 + {\cal S}_2.
\ee
The most probable value of $V_1$ is such that this entropy is maximum, under the constraint 
$V_1+V_2=V$. 
Therefore, one has:
\be
\frac{\partial{\cal S}_1}{\partial V_1} = \frac{\partial{\cal S}_2}{\partial V_2}.
\ee
In thermodynamics, this quantity (which is constant throughout the system) is equal to $P/T$, 
where $P$ is the
pressure. In the context of granular materials, $\partial{\cal S}/{\partial V}>0$ was called 
the
inverse compactivity by Edwards, and noted $1/X$.

A reasonable requirement for ${\cal S}(V)$ is that it should be proportional to the number of 
grains $N$, and 
only depends on the free-volume fraction $\Phi$: ${\cal S}(V)=N s(\Phi)$. One also expects that 
$s(\Phi)$
vanishes for $\Phi \to 0$. A possibility, suggested in \cite{FL,Lemaitre}, is that each grain 
has a number 
of possible positions proportional to the free-volume, which leads to 
$s(\Phi)=\log \Phi/\upsilon_0$, where
$\upsilon_0$ is again a `quantum' of free-volume. Using this form for $s(\Phi)$, we finally 
find:
\be
\frac{\partial{\cal S}}{\partial V} = \frac{1}{X}=\frac{N^2 \upsilon}{V^2} 
\frac{\partial{s}}{\partial \Phi}
\approx \frac{\rho^{*2} \upsilon}{\Phi} \qquad (\Phi \ll 1),
\ee
showing that the Edwards `temperature' vanishes for $\Phi \to 0$.\footnote{Note the following 
intriguing consequence of 
Edwards postulate.
Suppose that the two compartments contain grains of the very same material, but with different
sizes (volume) $\upsilon_1$
and $\upsilon_2$, such that  $N_1 \upsilon_1 = N_2 \upsilon_2$, i.e. the volume fraction occupied
by the grains in the
two compartments is the same. The equality $X_1=X_2$ then imposes that 
$\Phi_1/\Phi_2=\upsilon_2/\upsilon_1$, i.e.
that smaller grains will have a higher free volume. This conclusion holds for more general 
choices of the entropy $s(\Phi)$.}
Now, consider a small region immersed in a large container with grains, which acts as
a reservoir 
of free volume. Exactly as in thermodynamics, one can show that the probability that a 
free-volume equal to
$\rm v$ is found in this region is given by:
\be
p(\rm v) \propto \exp(-\frac{\rm v}{X}) 
\ee
The probability for a hole of the size of a grain $\upsilon$ to appear is therefore: $p(\rm v=\upsilon) 
\sim \exp(-\Phi^{*2}/\Phi)$,
where $\Phi^* \equiv \rho^* \upsilon$. Assuming that the grain motion takes place when a
sufficiently large hole appears, one finds that the rate of compaction has the same 
exponential form as above.

\subsection{Granular dynamics and the trap model}

The above `free-volume' ideas are interesting and certainly contain important physical 
ideas. However, the 
description of the dynamics using a single macroscopic degree of freedom (namely the 
average free volume density) is too
naive to account for the more sophisticated `memory' experiments. Indeed, any rate 
equation of the form:
\be
\frac{d\Phi}{dt} = -\gamma(\Phi,\Gamma) \Phi,
\ee
is unable to explain why a packing prepared under different tapping conditions 
(different $\Gamma$), but such as to
reach the same value $\Phi_0$ would evolve differently if tapped with the same amplitude.
The same argument was 
used by Kovacs in the context of glasses to suggest that additional parameters are needed to fully 
describe a glassy state. 
Take the example of a one dimensional ferromagnetic Ising model that relaxes towards its 
equilibrium state a 
temperature $T$. Since the system does not order, the equilibrium state is characterized by
a density of domain walls 
(or kinks), defining a characteristic domain size, which is the distance between the kinks. 
The non equilibrium state can also be 
characterized by a density of kinks, which decays with time towards the equilibrium. 
However, the full description 
of the non equilibrium state requires the specification of the {\it distribution} of 
the domain size -- only its
first moment is fixed by the average density of kinks. Different initial distributions of 
domain sizes with the same 
average value do evolve differently at a given temperature. For example, if there is an 
initial excess probability of large 
domains (but such that the density of kinks is the equilibrium one), these will immediately break down, leading to a temporary increase of the density 
of kinks.

A simple model that encapsulates both the spatial heterogeneity and the intermittency of 
the dynamics, is the `trap'
model. One should think of a glassy system as made of independent subsystems of a certain 
size $\xi$ (see Fig. \ref{blobfig}). Inside each 
of these regions, the dynamics is `coherent' in the sense that hopping between different 
metastable states involves all particles within a blob of size $\leq \xi$. Within each of these subunits, the dynamics can be thought of 
as a random walk in a rugged 
landscape, an idea with a long history in the context of glasses \cite{Goldstein}, 
and witnessing
a strong recent revival in
different contexts \cite{KL96,I01,Grigera}. However, an element which often seems to be missing 
from the discussion is the fact 
that the dynamics of system 
{\it as a whole} necessarily results (for short range interactions) 
from the evolution in parallel of many subsystems: the dynamics of a particle 
in a given subregion is completely unaffected by the dynamics of far away particles. 
An open problem 
is to identify precisely the 
(possibly time dependent) coherence length $\xi$, and understand its 
temperature and density 
dependence (see \cite{Heuer2} for very interesting results on this 
aspect in the context of Lennard-Jones systems). 

\bfig[hbt]
\bc
\epsfysize=8cm
\epsfbox{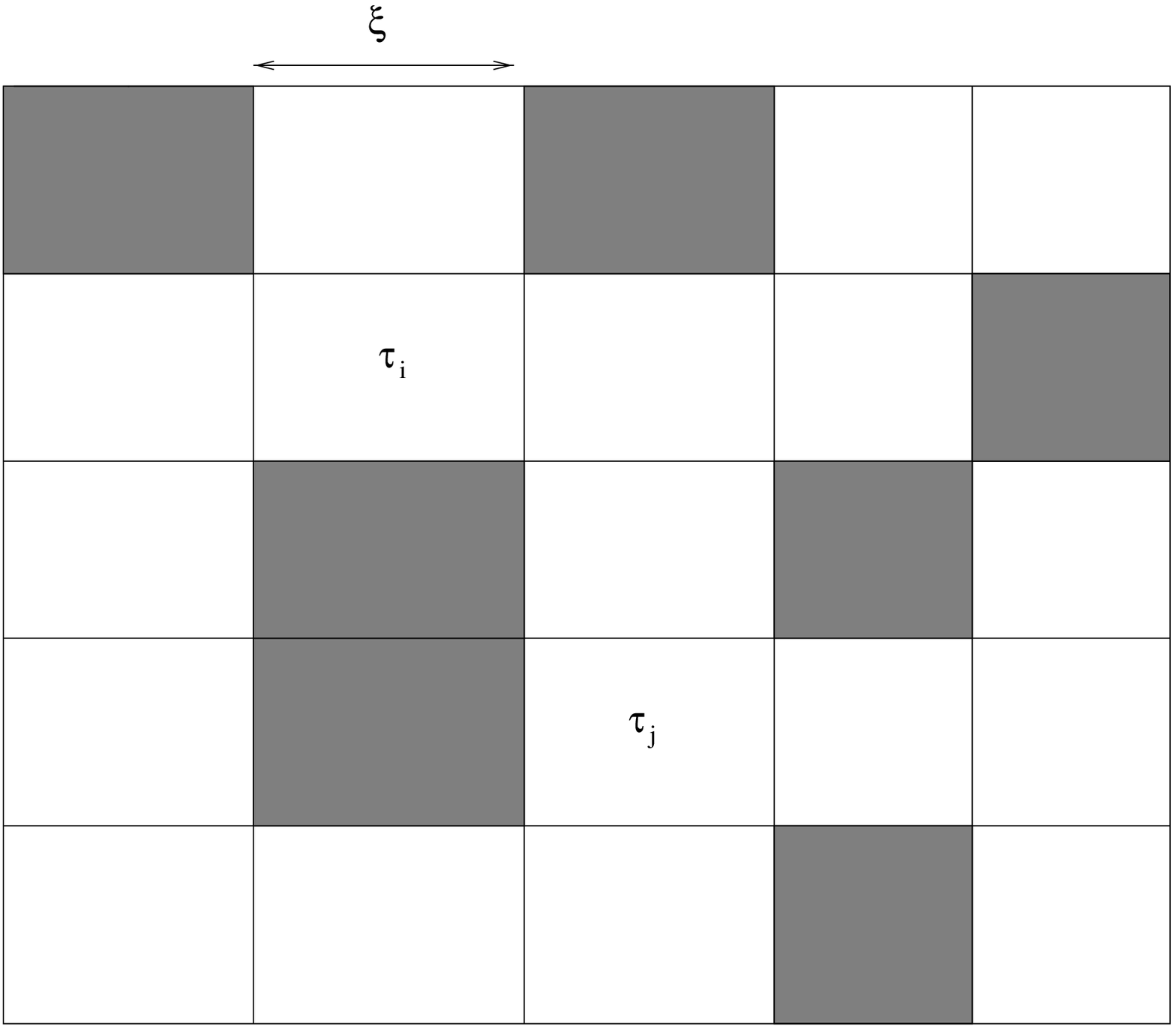}
\caption{\small Schematic view of the dynamics in glassy material. Within each `blob', the
dynamics is `coherent' in the sense that hopping between different 
metastable states involves all particles within a blob of size $\leq \xi$, whereas the
motion of far away particles does not affect the dynamics within this blob. 
Within each of these subunits, the dynamics can be thought of as a random walk in a rugged 
landscape, with an instantaneous (random) trapping time $\tau_i$ in the $i$th blob. Grey
regions are particularly `slow' ($\tau \gg t_w$). 
\label{blobfig}}
\ec
\efig

The dynamics of a given subsystem can be seen as a succession of hops between different
metastable states, each of
which blocking the dynamics for a time that depends on the local packing fraction: high 
densities leads to long 
trapping times. The state of the system at time $t$ is described by a {\it probability 
distribution} $P(\phi,t)$,
which counts the number of subsystems with a local value $\phi$ of the free volume. 
Each time a subunit unjams,
we assume that it falls back into any of the blocked states with an {\it a priori} 
distribution $P_0(\phi)$, that
reflects the number of states with a given packing density. The time evolution of 
$P(\phi,t)$ is then given by:
\be
\frac{d P(\phi,t)}{dt} = -\frac{1}{\tau(\phi)} P(\phi,t) + \left\langle 
\frac{1}{\tau(\phi)} \right\rangle_t P_0(\phi).
\ee
This equation was introduced in \cite{Dyre,BCM,MB} in the context of glasses, and 
extended to granular media by D. Head \cite{Head2}, 
where the free volume density $\phi$ plays the r\^ole of the energy $E$. 
This equation neglects any coupling between 
nearby sub-systems. This is
obviously unrealistic since the free volume liberated at one point will in general help nearby 
volumes to unjam. A generalization
of the above model that takes into account, to some extent, this coupling, can be 
found in \cite{BCM,MB,Head1}. 

The trap model, and its `SGR' (Soft Glassy Materials) generalisation covered in Mike Cates lectures
\cite{SGR,SGR2}, exhibits 
a number of interesting features, such as a 
genuine glass transition, aging and non linear rheology. A crucial ingredient of the 
trap model is the possibility 
for the average trapping time to diverge. Taking $\tau(\phi) \sim \exp(A/\phi)$, 
where $A$ is a certain constant that may depend on the tapping amplitude $\Gamma$, 
and $P_0(\phi)=\exp[\xi^3 s(\phi)]$ with $s(\phi)=B\log \phi$, we find that the 
distribution of trapping times $\Psi(\tau)$ 
decays very slowly, as $1/\tau (\log \tau)^a$, where $a=2+B\xi^3$. 
This means that one is always in the 
glassy phase of the trap model, since $\langle \tau \rangle = \infty$.\footnote{Note that
the logarithmic tail found above can be seen as an effective power-law with a time 
dependent effective exponent $\mu=a/\log \tau$. This suggests that the relevant coherence 
length after time $t$ is such that $B \xi^3 \sim \log t$.}
Correspondingly, 
one expects not only slow 
(logarithmic) compaction but also aging
effects, such as reported in numerical simulations \cite{Nicodemi}. In the glassy 
phase, the trap model describes the dynamics of each subsystem as essentially
{\it intermittent}: either the system is blocked, or it moves fast to a 
quite different configuration; most of the time is spent in one particularly
well jammed configuration. This intermittent dynamics of glassy systems begins to have
some numerical \cite{Heuer,Reichman,Berthiernew} and experimental 
\cite{Israeloff,Ciliberto,Cip} support. However, one should remember that if one
looks at the system as a whole, the activity will be dominated by the 
fastest regions (which can in turn, through the coupling between nearby 
that we have neglected above, unlock the slow regions). 

Interestingly, the trap model is able to reproduce some of the cycle effects 
reported above: the irreversibility in a
tapping cycle \cite{Head2}, or the Kovacs effect \cite{Bertininprep,Viasnoff}.
Variants of the trap model can 
also be studied, where each `trap' is decorated by the dynamics of smaller length 
scales \cite{BD,StA} in order
to account for memory and rejuvenation effects observed in spin-glasses and other
glassy systems. The Sinai model
is in this family of `multi-scale' trap models \cite{LeD,LeD2,Sales}, and was introduced 
in the context of granular 
media precisely to understand the memory and cycle effects \cite{Pouliquen}, although 
the simpler `monoscale'
trap model seems to be able to account for most of them. It would be very interesting
to exhibit experimentally a
truly `multiscale' dynamical phenomenon in granular materials. This would have the obvious 
advantage over many other glassy systems that the underlying mechanism, in terms of 
embedded length scales, can be directly observable \cite{StA,PRB}.

There would be much more to say about glassy dynamics in the context of granular materials, 
but my lectures stopped at this point, after only quite simple ideas were expressed.  
Since some of the more elaborated concepts are common to spin-glass dynamics and glassy 
rheology, it is appropriate to refer to the
lectures of Mike Cates, Leticia Cugliandolo and Giorgio Parisi in this volume, and also
to \cite{BerthierHouches,Jamming,Review,Berthier-Iguain} for further developments.

\end{document}